\documentclass[traditabstract]{aa}

\usepackage{graphicx}
\usepackage{txfonts}
\usepackage{amsmath}
\usepackage{subcaption}
\usepackage{booktabs}
\usepackage[hidelinks]{hyperref}

\begin{document}

\title{Cross-simulator transfer with foundation-model summaries:  \\ Toward robust SKA-era reionization inference}

\titlerunning{Cross-simulator transfer with foundation-model summaries}

\author{Y.~Pietschke\inst{\ref{inst1}}\thanks{e-mail: pietschke@thphys.uni-heidelberg.de}
        \and
        C.~Heneka\inst{\ref{inst1}}
        \and
        A.~Ore\inst{\ref{inst1}}
        \and
        R.~Meriot\inst{\ref{inst2}}
        }

    \authorrunning{Pietschke et al.}

    \institute{Institut f\"ur Theoretische Physik, Universit\"at Heidelberg, Philosophenweg 16, 69120 Heidelberg, Germany
   \label{inst1}
   \and
   Max Planck Institute for Radioastronomy, Auf dem H\"ugel 69, Bonn, 53121, Germany \label{inst2}
         }

\date{Received <date>; accepted <date>}
\abstract{
Simulation-based inference (SBI) for parameter estimation is vulnerable to model misspecification: neural summaries and density estimators trained on a specific forward model typically fail when applied to data drawn from another model or from real observations, and no training simulator can capture the full observational pipeline of a real measurement exactly. We show that a self-supervised vision transformer (ViT), pretrained label-free on a fast approximate simulator, produces transferable data summaries that generalize across simulators. Without retraining, it can be reused as a frozen encoder to infer astrophysical parameters from a completely different simulator that explicitly resolves radiative transfer and on which it was never trained. As a concrete use case in 21cm cosmology, SKATR, a ViT pretrained with a self-supervised Joint Embedding Predictive Architecture (JEPA), serves as a foundation model for reionization inference from upcoming Square Kilometre Array (SKA) measurements: SKATR was pretrained once on 67k low-cost, noiseless semi-numerical 21cmFAST light cones, frozen, and applied to hydrodynamical Loreli~II light cones, where a lightweight conditional flow matching (CFM) head infers five astrophysical parameters. The encoder is never shown Loreli data, its parameters, or any noise.
In our comparison, SKATR yields the most precise and best-calibrated posteriors across all five parameters, matching the accuracy of the fully supervised in-domain baseline while requiring $2.6\times$ fewer radiative-transfer simulations. Under realistic SKA~AA$^\ast$ noise, only SKATR remains simultaneously accurate, informative, and calibrated, outperforming even a supervised baseline retrained from scratch on noisy data. Self-supervised pretraining on computationally efficient semi-numerical simulations is therefore a viable route to calibrated, simulator-, and noise-agnostic reionization inference for the SKA-era.}


\maketitle
\nolinenumbers

\section{Introduction}
\label{sec:intro}

Simulation-based inference (SBI)~\citep{Cranmer_2020} has emerged as a valuable tool for cosmological parameter estimation from data whose likelihood is intractable, circumventing explicit likelihood modeling by learning the map from observation to parameters directly from forward-simulated data~\citep{Alsing_2019, Cole_2022, Saxena_2024, Schosser_2025}. Simulation-based inference is a flexible framework for inference from either hand-crafted summary statistics, such as the power spectrum, network-learned summaries, or field-level inference, as well as combinations thereof in terms of hybrid statistics~\citep{Zhao_2022a, Makinen_2024, Prelogovic_2024, Schosser_2026}.  Its central bottleneck, however, is that neural summaries and density estimators trained on simulated data typically fail when applied to observations, or even to data from a different simulator: distinct codes adopt distinct parameterizations and physical approximations, and no single simulator can precisely capture the full instrumental and observational pipeline of a real measurement~\citep{Cannon_2022, Sooknunan_2024}. This model-misspecification problem is the fundamental barrier between forthcoming survey data and calibrated cosmological constraints, and it is not unique to any one probe.

A route that has recently emerged to address such distribution shifts is self-supervised foundation-model pretraining. The idea, established for language and vision and increasingly explored in astronomy \citep[e.g.,][]{Parker_2024, Parker_2025, Smith_2024}, is to pretrain a large encoder once on a low-cost and abundant dataset with a label-free objective, then reuse the frozen representation for many downstream tasks with light task-specific supervision. Because the pretraining objective is not tied to any single target label set, the learned representation can, in principle, transfer to targets whose parameters, noise model, and systematics were never seen during pretraining. This is precisely the regime where model misspecification typically produces biased or miscalibrated results. An established alternative for settings with known target data is domain adaptation, which learns domain-invariant features by aligning a labeled source to a specific target domain~\citep{Roncoli_2023, Ntampaka_2025, Pierre_2026}. As this approach requires access to the target domain, it is not available for genuinely unseen future observations. Target-agnostic, self-supervised representations bypass this constraint.

In this work we demonstrate foundation-model pretraining in the context of inference for 21cm cosmology, which is one of the main science cases of the Square Kilometre Array (SKA)~\footnote{https://www.skao.int/en}. SKA radio interferometric observations at low frequencies of $50-350\,$MHz will illuminate the epoch of reionization (EoR) and cosmic dawn -- key milestones in the history of the Universe, spanning redshifts from roughly 5 to 20. Here, the first luminous sources formed and subsequently reionized the surrounding intergalactic medium (IGM), representing the Universe's last phase transition from neutral to ionized. Mapping the redshifted 21cm hyperfine transition of neutral hydrogen (HI) offers a window into these epochs, enabling studies of the IGM and providing 3D tomographic light cones of the early Universe. Precursor experiments, such as the Murchison Widefield Array (MWA), the Hydrogen Epoch of Reionization Array (HERA), and the LOw-Frequency ARray (LOFAR), have set progressively tighter upper limits on the 21cm power spectrum~\citep{ Parsons_2010, Tingay_2013, DeBoer_2017, LOFAR, Mertens_2020, Trott_2020, Yoshiura_2021, Abdurashidova_2022, Abdurashidova_2023, Ceccotti_2025}, while SKA will uniquely enable imaging of the reionization signal at the sensitivity required for calibrated parameter inference. These large-scale surveys of 21cm line fluctuations allow us to constrain the astrophysics of galaxies and the IGM~\citep{Park_2019}, as well as probe cosmology and fundamental physics~\citep{Liu_2020}.

Model misspecification for 21cm cosmology has two aspects. First, any SKA measurement will carry thermal noise, ionospheric effects, foreground residuals, as well as beam and calibration artifacts that no training simulation can precisely capture, creating a domain shift between the source and target data. A second gap can be observed between different forward models. Distinct simulators use distinct parameterizations and physical approximations, and current 21cm EoR models can differ by at least $10\%$~\citep{Zahn_2011, Greig_2018_21cmmc}, which biases inference. For SKA we do not know in advance which simulator will best match future observations. This fragility shows up not only in posterior accuracy and precision but, crucially, in calibration, namely during correct posterior coverage. Both a precise-but-inaccurate posterior (concentrated on the wrong answer) and an accurate-but-overconfident one (recovering the right mean but underestimating its own uncertainty) would silently mislead the analysis.

Target-agnostic representations can circumvent the model misspecification problem by learning summary statistics tied to the physics of reionization itself rather than to the parameters of a single simulator or noise model. Self-supervised representation learning, in particular Joint Embedding Predictive Architectures (JEPA)~\citep{Assran_2023, Bardes_2024}, is a promising candidate: the encoder is trained to predict masked regions of the input in embedding space, without any label. \citet{Ore_2025} introduced the SKA Transformer (SKATR), a vision transformer (ViT)~\citep{Dosovitskiy_2021} pretrained with JEPA on 21cmFAST light cones. They show that the resulting summaries are robust to additions of optimistic SKA-Low (AA4) noise, to changes in spatial resolution, and to changes in the set of simulation parameters within a single simulator family. The observationally pressing question is whether such summaries transfer across simulator families, and thus parameterizations, and across noise regimes at once. If they do, the same frozen encoder can be used to perform calibrated inference on data whose physics and systematics were never seen during pretraining.

In this work we test exactly this: whether SKATR enables transfer across simulator models. Our aim is to show that a label-free, symmetry-agnostic pretraining objective yields transferable summaries, without hand-designed augmentations built into the foundation model. SKATR was pretrained once on a large set of low-cost, noiseless semi-numerical light cones, frozen, and applied to light cones from a different and more expensive hydrodynamical simulator (first noiseless, then with realistic SKA-Low noise in the AA$^\ast$ array configuration). A lightweight conditional flow matching (CFM) inference head, which is an updated version of the EoRFlow pipeline of \citet{Pietschke_2025}, was trained on the resulting summaries to recover the target astrophysical parameters. The encoder is never exposed to the target hydrodynamical simulator's data, its parameters, or the noise. This is the standard foundation-model recipe, pretrained once on large data (here, approximate semi-numerical simulations), transferred with light task-specific supervision, applied for the first time successfully to 21cm inference under a joint simulator-and-noise domain shift.

This paper is structured as follows. We begin with a description of our databases, highlighting the differences in physical modeling of the EoR for our two different simulators in Sect.~\ref{sec:sims}. Moving on, we summarize the SKATR network architecture and our experimental setup in Sect.~\ref{sec:network}. In Sect.~\ref{sec:results_clean}, we present our results for cross-simulator transfer on noiseless simulations. In Sect.~\ref{sec:results_noise} we discuss adding realistic instrument noise to provide a robust inference pipeline for SKA data. Finally, we discuss our findings and outline directions for future work in Sect.~\ref{sec:conclusion}.

\section{Simulation methods}
\label{sec:sims}
This work uses two distinct simulators of the 21cm signal that differ in physics, in parameterization, and in compute cost. Our source data, which SKATR is trained on, are based on 21cm light cones generated with the semi-numerical code \texttt{21cmFAST}~\citep{21cmfast11}. With light cones generated comparably fast, they form the larger label-free training set for self-supervised pretraining. The target data are hydrodynamical, radiative-transfer~(RT) light cones from the Loreli~II database, generated with \texttt{Licorice}. The ionization and thermal history follows from explicit RT rather than a subgrid prescription, but each simulation is two orders of magnitude more expensive per light cone and consequently the feasible database size is smaller. For the two simulators we also varied different sets of parameters in terms of astrophysics and number of parameters (six in the \texttt{21cmFAST} database we generated, five in the Loreli~II database), so transferring from one to the other crosses physics, parameterization, and (later) noise recipes at once. Our goal is to test how well self-supervised summaries learned from the lower-cost source data transfer to a more expensive but data-restricted simulator.

\subsection{Semi-numerical source data: 21cmFAST light cones}
\label{sec:21cmfast}
Our source dataset comprises $67{,}325$ semi-numerical light cones generated with \texttt{21cmFASTv4}~\citep{Park_2019,21cmfast}. 21cmFAST produces a discretized 3D field of the 21cm brightness-temperature offset $\delta T_b(\mathbf{x},z)$ as a function of on-sky position $\mathbf{x}$ and redshift (frequency) $z$. Initial density and velocity fields were generated in Lagrangian space and evolved with first- and approximate second-order Lagrangian perturbation theory. Ionized regions were then identified with the excursion-set approach, applying a real-space top-hat filter on decreasing scales and flagging a cell as ionized where the cumulative number of ionizing photons produced by ionizing sources (galaxies) exceeds the number of recombinations. We retained the full calculation of the spin temperature $T_\mathrm{S}$ rather than assuming postheating with $T_\mathrm{S} \gg T_\mathrm{CMB}$, so that X-ray heating and Lyman-$\alpha$ (Wouthuysen-Field) coupling are followed self-consistently. This is essential over the wide redshift range we simulate, which reaches into the cosmic dawn at $z\gtrsim 10$.
The brightness-temperature offset follows the usual form

\begin{equation}
\begin{split}
\delta T_\mathrm{b}(\mathbf{x},z) \simeq{}& 27\, x_\mathrm{HI}(\mathbf{x})\,
  \big(1+\delta_\mathrm{b}(\mathbf{x})\big)
  \left(1-\frac{T_\mathrm{CMB}}{T_\mathrm{S}}\right)\\
  &\times \left(\frac{1+z}{10}\,\frac{0.15}{\Omega_\mathrm{m} h^2}\right)^{1/2}
  \left(\frac{\Omega_\mathrm{b} h^2}{0.023}\right)\\
  &\times \left(\frac{H}{\mathrm{d}v_r/\mathrm{d}r + H}\right)\,\mathrm{mK},
\end{split}
\label{eq:dTb}
\end{equation}

with $x_\mathrm{HI}$ the neutral-hydrogen fraction, $\delta_\mathrm{b}$ the baryon
overdensity, $T_\mathrm{S}$ the spin temperature, and $\mathrm{d}v_r/\mathrm{d}r$
the line-of-sight peculiar-velocity gradient~\citep{21cmfast11}.

We varied six astrophysical parameters of the galaxy source model
of~\citet{Park_2019}. The stellar mass fraction and the ionizing escape
fraction are

\begin{align}
f_\star(M_h) &= f_{\star,10}\left(\frac{M_h}{10^{10}\,M_\odot}\right)^{\alpha_\star}, \\
f_\mathrm{esc}(M_h) &= f_\mathrm{esc,10}\left(\frac{M_h}{10^{10}\,M_\odot}\right)^{\alpha_\mathrm{esc}},
\end{align}

where $f_{\star,10}$ and $f_\mathrm{esc,10}$ are the normalizations at a
halo mass of $10^{10}\,M_\odot$ and $\alpha_\star$, $\alpha_\mathrm{esc}$
the corresponding slopes. The X-ray heating of the intergalactic
medium is controlled by the specific X-ray luminosity $L_\mathrm{X}$ (the soft-band
X-ray luminosity per unit star-formation rate that escapes host galaxies)
and the energy threshold $E_0$ below which X-rays are self-absorbed in the
host galaxy and do not contribute to heating. The remaining galaxy
parameters (such as the minimum turn-over halo mass for star formation
and the star-formation timescale) were held fixed at their fiducial
values~\citep{Park_2019}, and the cosmology is fixed to Planck
values~\citep{Planck2018}. Parameter values were drawn from flat priors
over the ranges in Table~\ref{tab:source_params}. In addition to these
astrophysical parameters, we also varied the random seed to account for
cosmic variance across the dataset.

\begin{table}[h]
\centering
\small
\caption{Varied astrophysical parameters of the source (21cmFAST)
dataset and their flat prior ranges.}
\label{tab:source_params}
\begin{tabular}{ll}
\toprule
Parameter & Prior \\
\midrule
$\log_{10} f_{\star,10}$      &  [-3, 0] \\
$\alpha_\star$                & [-0.3, 0.9] \\
$\log_{10} f_\mathrm{esc,10}$ & [-3, 0] \\
$\alpha_\mathrm{esc}$         & [-0.8, 0.5] \\
$\log_{10} L_\mathrm{X}$ [erg\,s$^{-1}$]              & [38, 42] \\
$E_0$ [eV]                        & [100, 1500] \\
\bottomrule
\end{tabular}
\end{table}

Each light cone was simulated in a $200\,\mathrm{cMpc}$ box on a
$160\times160\times1920$ voxel grid, giving a transverse resolution of
$1.25\,\mathrm{Mpc}$, and spans the redshift range $z = 5$--$35$ along the
line of sight. With $\sim$15\,min ($\sim\!0.25$\,CPU-h) per light cone and $\sim 67$k light cones generated, this
large and comparably lightweight source set is well suited to the label-free pretraining of
SKATR.

\subsection{Hydrodynamical target data: Loreli~II light cones}
\label{sec:loreli}
The target data consist of $7{,}667$ Loreli~II light cones~\citep{Meriot_24,Meriot_2025} generated with the hydrodynamical RT code \texttt{Licorice}~\citep{Semelin_2007, Baek_2009, Semelin_2016, Semelin_2017}. \texttt{Licorice} is a combined $N$-body~and~SPH code that performs full 3D RT. Star-forming gas particles emit UV and X-ray photon packets that propagate on an adaptive grid and update the temperature and ionization state of the gas. In contrast to the semi-numerical excursion-set treatment of the source dataset, the ionization and thermal histories here follow from explicit RT. This is a fundamental difference in physical modeling that makes transfer between the two simulators a nontrivial test of representational generalization. The database is run at a moderate resolution of $256^3$ resolution elements in a $\sim300\,\mathrm{Mpc}$ box.  At this resolution \texttt{Licorice} employs a conditional-mass-function subgrid model for star formation in unresolved halos and a two-phase treatment where both the average temperature and the temperature of the
neutral phase of the particles are computed separately. The latter is essential to avoid biasing the 21cm signal at this resolution~\citep{Meriot_24}. The cosmology is fixed to Planck~2018~\citep{Planck2018}, as is our semi-numerical database described in the previous section.
The 21cm brightness-temperature offset $\delta T_b$ was computed in post-processing following Eq.~\ref{eq:dTb}. The particle properties were interpolated onto a grid for the neutral-hydrogen fraction, baryon overdensity, peculiar-velocity gradient, and gas temperature; the Wouthuysen-Field coupling was evaluated with the semi-analytic code \texttt{SPINTER}~\citep{Semelin_2023} for the spin temperature, and $\delta T_b$ follows at each redshift. The original comoving boxes were calculated on a grid of $256^3$ voxels. For this work we use light cones downsampled by a factor of 2 in the transverse plane to a $128\times128\times1536$ voxel grid ($\approx\!2.3\,\mathrm{Mpc}$ per cell). All other simulation details follow~\citet{Meriot_24} and~\citet{Meriot_2025}.

Loreli~II varies five astrophysical parameters, shown in Table~\ref{tab:target_params}, sampled on a structured grid rather than from continuous priors. Star formation is controlled by the gas-to-star conversion timescale $\tau$ and the minimum halo mass for efficient star formation $M_\mathrm{min}$, jointly log-sampled as 42 pairs. The ionizing output is set by the late-time escape fraction $f_\mathrm{esc,post}$, which takes the three discrete values $\{0.05,\,0.275,\,0.5\}$ for particles in cells where the spatially averaged ionization fraction exceeds the threshold $\langle x_\mathrm{HII}\rangle > 0.03$; below that threshold $f_\mathrm{esc,post}$ is set to $0.003$. X-ray heating is determined by the X-ray production efficiency $f_X$ (13 logarithmically spaced values), which scales the soft-band X-ray luminosity per unit star-formation rate, and the hard-to-soft X-ray ratio $r_\mathrm{H}$ (six linearly spaced values).

\begin{table}[h]
\centering
\small
\caption{Varied astrophysical parameters of the target (Loreli~II) dataset.}
\label{tab:target_params}
\begin{tabular}{@{}l p{0.40\columnwidth} l@{}}
\toprule
Parameter & Meaning & Grid \\
\midrule
$\tau$ [10 Myr]        & star-formation timescale       & $[700,\,10\,500]$ \\
$M_\mathrm{min}$ [$\mathrm{M_\odot}$]          & min.\ star-forming halo mass   & $[10^{8},\,4\times10^{9}]$ \\
$f_X$                      & X-ray production efficiency    & $[0.1,\,10]$ \\
$r_\mathrm{H}$                      & hard-to-soft X-ray ratio       & $[0,\,1]$ \\
$f_\mathrm{esc,post}$      & late-time escape fraction      & $\{0.05,\,0.275,\,0.5\}$ \\
\bottomrule
\end{tabular}
\tablefoot{
$M_\mathrm{min}$ sampled jointly with
$\tau$ as 42 pairs;
$f_X$ sampled logarithmically over 13 values;
$r_\mathrm{H}$ sampled linearly over six values~\citep{Meriot_2025}.}
\end{table}

Loreli~II covers the redshift range $z = 5$--$15$, with reionization completing (volume-averaged $x_\mathrm{HII}\sim1$) at redshifts between $z \approx 5$ and $z \approx 8$ across the parameter grid. It is calibrated with star-formation observations of~\citet{Bouwens_2016,Oesch_2018,McLeod_2016} and produces Thomson optical depths within $3\sigma$ of the Planck~2018 value. Loreli~II is therefore consistent with current observational constraints, which makes it a well-motivated target for SKA-era inference. A single complete Loreli~II 21cm light cone at the $128\times128\times1536$ resolution we use, including the RT hydrodynamics and the subsequent 21cm post-processing, requires $300$--$500$\,CPU-h.

\subsection{Dataset comparison and processing}
\label{sec:data}
The compute-cost asymmetry between the two simulator families directly motivates our approach of cross-simulator transfer for inference. The $67{,}325$-light cone 21cmFAST training set costs $\sim\!1.7\times10^4$\,CPU-h compared to $\sim\!3.1\times10^6$\,CPU-h for the full $7{,}667$-light cone Loreli~II suite. This represents a $\sim\!180\times$ reduction in cost for a $\sim\!8.8\times$ larger set. Note that this comparison is not made at matched simulation resolution. Matching the Loreli~II native grid of $256^3$ would raise the 21cmFAST source data cost, but not by enough to change the order of magnitude of the gap.
The foundation-model strategy we adopted is designed to exploit precisely this asymmetry: it concentrates the heavy lifting on the approximate simulator and reserves the expensive simulator for a lightweight, target-specific inference head.

The 21cmFAST light cones we produced cover a redshift range of $z = 5$--$35$, while Loreli~II (in the version used here) only reaches up to $z\sim15$. To ensure SKATR gains no advantage from the larger temporal range during pretraining, we crop the 21cmFAST light cones to the redshift interval covered by Loreli~II, while keeping the line-of-sight length required by the ViT patch grid (Sect.~\ref{sec:network}). This results in the common redshift range $z = 5$--$13.15$. This redshift mismatch introduces an additional out-of-distribution challenge for SKATR, whose pretraining light cones do not cover the full temporal range of the target data. A supervised baseline, by contrast, would need retraining on light cones matching the Loreli redshift coverage exactly. Both datasets are then downsampled to a $32\times32\times576$ voxel grid: the transverse plane is reduced to $32\times32$ by block averaging (a factor of $5$ for the source and $4$ for the target according to their native resolutions) and the line of sight is averaged to $576$ cells. Finally, the per-sample median brightness temperature is shifted to a common offset and a symmetric log transform $\log(|T|+1)\cdot\mathrm{sign}(T)$ compresses the dynamic range while preserving the sign. The shift value used for this transformation is computed in advance on each input dataset (noiseless 21cmFAST, noiseless Loreli, noisy Loreli) separately, ensuring that all three  are centered on the same value before encoding.

To prevent data leakage between encoder training, inference-head training, and evaluation, we fixed one canonical Loreli test split, reused across pipelines and training budgets. Of the $7{,}667$ target light cones, $1{,}150$ were held out as a test set never seen during any training stage. The remaining $6{,}517$ form the pool from which training sets are drawn. We swept the training budget $N \in \{100, 500, 1000, 2500, 5000, 6517\}$ by subsampling this pool, always evaluating on the same held-out test set.

\subsection{SKA AA$^\ast$ noise model}
\label{sec:noise}
To ensure our transfer approach remains robust when applied to SKA data,
we added a second domain shift on top of the cross-simulator one by
corrupting the Loreli target
light cones with realistic SKA AA$^\ast$ noise and applying a foreground-avoidance cut. Compared to the
noise transfer study in the original SKATR paper~\citep{Ore_2025}, the setup is
substantially more conservative. The array configuration changes
from the full AA4 design to the operational SKA1-Low survey
assembly AA$^\ast$, which is the configuration that more near-term SKA
reionization measurement will use. The dominant difference is
in the foreground treatment: \citet{Ore_2025} adopted a more optimistic
assumption where the foreground
wedge extends only up to the primary field of view of the instrument. Instead, we took the more
conservative, moderate approach where the foreground wedge extends $0.1\,h\,\mathrm{Mpc}^{-1}$ beyond
the horizon limit.

Per-mode sensitivities were computed for the SKA AA$^\ast$ array configuration with \texttt{21cmSense}~\citep{21cmsense13,21cmsense14}, sampling the relevant frequency range and assuming $1080$\,h of total integration time. The noise was injected on the post-coarsening $128\times128\times1536$ Loreli grid, prior to the common $32\times32\times576$ downsampling of Sect.~\ref{sec:data}, with a single noise realization drawn deterministically per light cone so that each light cone carries one fixed noise realization. Foreground avoidance was implemented in Fourier space by excising all modes inside the foreground wedge: the wedge edge was taken at $0.1\,h\,\mathrm{Mpc}^{-1}$ beyond the horizon limit, following the standard practice of current power spectrum analyses. After noise injection and wedge removal the per-sample voxel median is essentially zero, and the common offset described above was set accordingly. Crucially, noise only enters the target Loreli data; the SKATR and ViT encoders of 21cmFAST data remain those pretrained on noiseless 21cmFAST light cones, so their use on noisy data is a genuine zero-shot transfer across the
noise boundary.

\section{Network architecture and experimental setup}
\label{sec:network}
\subsection{SKATR encoder} \label{sec:skatr}
The SKA Transformer (SKATR)~\citep{Ore_2025} is a vision transformer~\citep{Dosovitskiy_2021}
adapted to 3D volumetric inputs and trained in a self-supervised fashion to learn compressed data representations. Each input light cone of shape $32\times32\times576$ (Sect.~\ref{sec:data}) is tokenized into nonoverlapping $[4, 4, 12]$ patches yielding $8\times8\times48 = 3072$ tokens per light cone.
The patches are augmented with learnable Fourier sin/cos positional encodings and linearly projected to a 360D token embedding.
The encoder stacks six pre-norm transformer blocks with six attention heads and a multilayer perceptron (MLP) hidden dimension of 720 ($\sim$6.3\,M parameters). The architecture follows~\citet{Ore_2025} apart from the patch shape, which we change from $[4, 4, 10]$ to $[4, 4, 12]$ to match the longer redshift axis of our input grid.

The encoder was pretrained with a Joint Embedding Predictive Architecture (JEPA)~\citep{Assran_2023}, which learns representations by predicting the embeddings of masked regions of the input rather than reconstructing them in voxel space, and requires no per-sample labels. The setup uses two ViTs of identical architecture, a context encoder and a target encoder. During training the target encoder embeds the full, patched light cone, while the context encoder sees only the unmasked (context) patches, with the target patches withheld from its input. A small predictor transformer (four blocks, four attention heads, hidden dimension 48, fixed sin/cos positional encoding) then takes the context embeddings together with mask tokens at the target locations and predicts the target-encoder embeddings there. The loss is the mean absolute error (L1) between the predicted and target embeddings at the masked positions. Computing this loss in embedding space rather than reconstructing voxels is a deliberate design choice, inherited from \citet{Ore_2025}. Within the machine-learning literature the two targets have been compared directly and repeatedly favor the embedding-space variant whenever the encoder is subsequently used frozen for inference-type and probing tasks. Ablating only the target space, \citet{Assran_2023} report a drop from $66.9\%$ to $40.7\%$ top-1 in low-shot ImageNet linear probing when an otherwise identical model predicts pixels instead of embeddings, and \citet{Bardes_2024} find feature targets to outperform normalized-pixel targets of the type used by masked autoencoders \citep{He_2022} on video, by $73.7\%$ against $68.6\%$ on Kinetics-400 under frozen evaluation. \citet{Balestriero_2024} offer a complementary argument, showing that reconstruction in input space preferentially allocates model capacity to the highest-variance components of the data, which need not be the components that matter downstream. We therefore take joint embedding-space prediction as the method of choice for transfer in a parameter inference task. We here test whether the resulting summaries transfer across simulators and noise models.

Target regions are sampled at mixed spatial scales to capture both bubble-scale and field-scale structure: each pretraining step draws four rectangular target regions per light cone -- two large ($60$--$80\%$ of the spatial area) and two small ($10$--$20\%$) -- each spanning the full redshift axis, so that masking acts purely in the sky plane and never along the line of sight. Each target region is a single rectangle and is predicted from the complement of that rectangle as its context.

Our training differs from the canonical JEPA prescription of~\citet{Assran_2023} in two aspects. First, where canonical JEPA conditions on a union of multiple context blocks, we predict each target rectangle independently from its own complementary context. Second, and more importantly, the target encoder is not an exponential moving average (EMA) of the context encoder but a copy of it at initialization, held fixed throughout training. A fixed target cannot collapse, so the EMA method that canonical JEPA relies on to stabilize training and prevent representational collapse is not necessary here. The context--predictor path simply learns to reproduce the embeddings of a fixed, randomly initialized network of the same architecture, which are known to be informative targets for self-supervised learning~\citep{Sui_2024}. The target encoder serves only to generate prediction targets during pretraining and is discarded afterward, and the summary network carried forward is the trained context encoder, not the random network. We find this fixed-target variant to be more stable during training than the EMA update in our setup. In our runs with an EMA target the validation loss reaches a minimum within the first ten to twenty epochs and then rises steadily, ending at up to $1.6$ times its minimum value, whereas the fixed-target runs decrease monotonically and remain flat once converged. We do not take this as evidence that an EMA target is inferior in general. A different architecture, or further tuning of the momentum schedule and learning rate, may well stabilize it, and we leave that exploration to future work.

Pretraining runs for 100 epochs on the $67{,}325$ 21cmFAST light cones with AdamW (peak learning rate $10^{-3}$ on a \texttt{OneCycle} schedule, batch size 64, gradient-norm clipping at 10). No Loreli data and no parameter labels are ever shown to the encoder, and pretraining takes $\sim$110\,GPU-hours on a single Nvidia A100, or $\sim$60\,GPU-hours on a single Nvidia H200. The encoder is then frozen at its best-validation epoch, which for this run is epoch 90 of 100. Training and validation loss curves for all encoders are shown in Appendix~\ref{app:losses}. We summarize a Loreli light cone by mean-pooling its 3072 context-encoder token embeddings into a single 360D vector, and this same frozen encoder is reused at every downstream Loreli training budget $N$
(Sect.~\ref{sec:baselines}).

\subsection{Inference head}
\label{sec:infhead}

\begin{figure*}
    \centering
    \includegraphics[width=\linewidth]{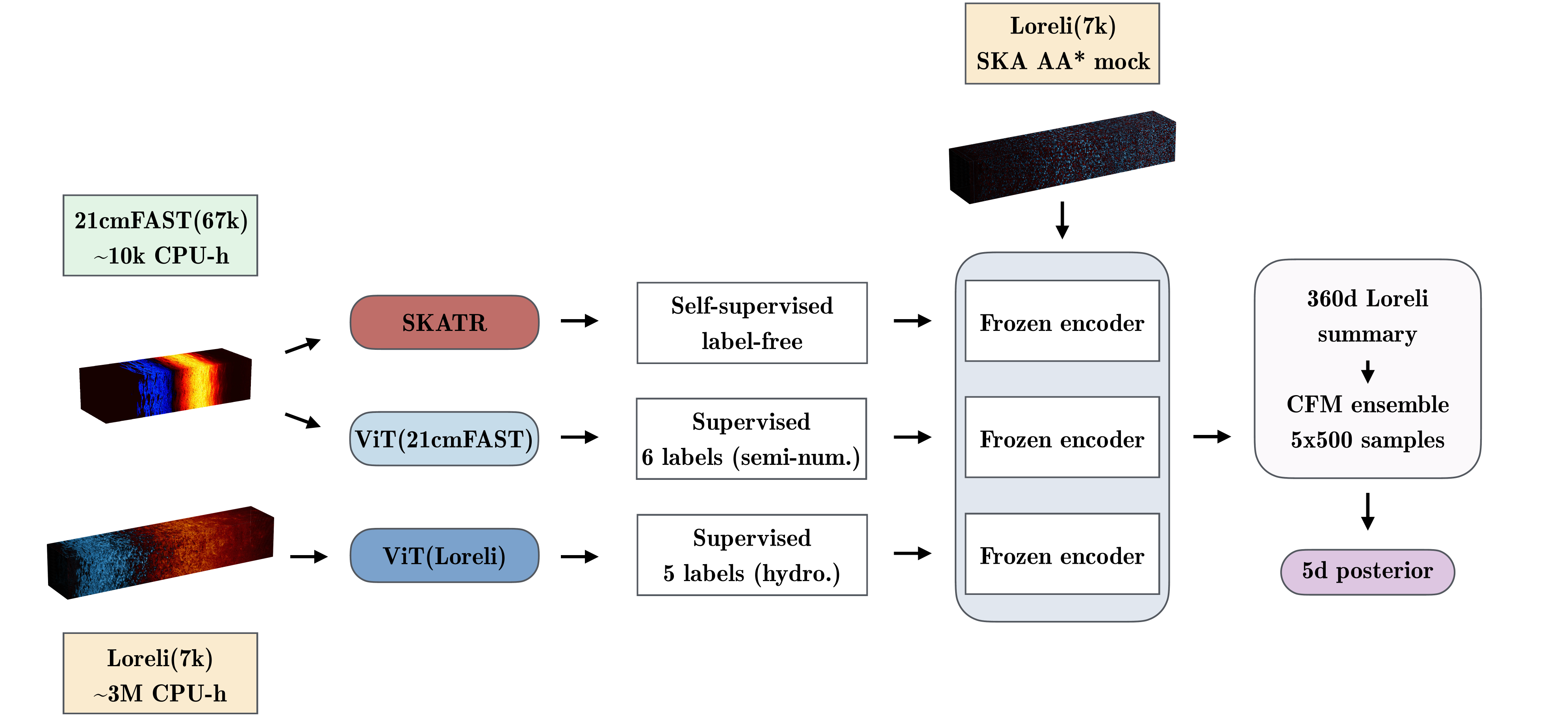}
    \caption{Three inference pipelines compared in this work. They share an identical downstream protocol: the frozen encoder maps a Loreli~II light cone to a 360D summary, which an ensemble of five CFM heads converts into a posterior over the Loreli astrophysical parameters. The pipelines differ only in how the encoder is pretrained. SKATR is pretrained and self-supervised on 67{,}325 semi-numerical 21cmFAST light cones with no parameter labels; ViT(21cmFAST) uses the same low-cost data but performs supervised regression on the six 21cmFAST parameters; ViT(Loreli) is trained and supervised directly on $N$ expensive Loreli~II light cones, retrained for every training budget of $N \in \{100, 500, 1000, 2500, 5000, 6517\}$ light cones.}
\label{fig:pipelines}
\end{figure*}

We mapped a frozen 360D summary to a posterior over the five Loreli astrophysical parameters with a lightweight CFM~\citep{Lipman_2023, Tong_2023} head, an updated version of the EoRFlow inference pipeline of~\citet{Pietschke_2025}. Conditional flow matching learns a continuous transport from a simple base density $q_0(\boldsymbol{\theta}_0) = \mathcal{N}(\mathbf{0}, \mathbf{I})$ to the target posterior $q_1(\boldsymbol{\theta}_1 | s) \approx p(\boldsymbol{\theta} | s)$ conditioned on the summary $s$, parameterized by a continuous time variable $t \in [0,1]$ along which a learned vector (velocity) field $v_\phi(\boldsymbol{\theta}_t, t, s)$ drives the flow $\mathrm{d}\boldsymbol{\theta}_t / \mathrm{d}t = v_\phi(\boldsymbol{\theta}_t, t, s)$. Following the sample-conditional, optimal-transport formulation of~\citet{Lipman_2023}, a parameter sample $\tilde{\boldsymbol{\theta}}$ is drawn from the joint distribution $p(\boldsymbol{\theta})\,\mathcal{L}(s|\boldsymbol{\theta})$ of prior and likelihood via the simulator, a base sample $\boldsymbol{\theta}_0 \sim \mathcal{N}(\mathbf{0}, \mathbf{I})$ is drawn independently, and the two samples are linearly interpolated at a uniformly sampled time,

\begin{equation}
\boldsymbol{\theta}_t = (1-t)\,\boldsymbol{\theta}_0
                      + t\,\tilde{\boldsymbol{\theta}},
\qquad t \sim \mathcal{U}(0,1).
\label{eq:cfm-interpolant}
\end{equation}

The corresponding true velocity field is $u_t =
\tilde{\boldsymbol{\theta}} - \boldsymbol{\theta}_0$, and the network
is trained to regress it with the mean-squared error:

\begin{equation}
\mathcal{L}_\mathrm{CFM}
   = \mathbb{E}_{\,t,\,s,\,\tilde{\boldsymbol{\theta}},\,\boldsymbol{\theta}_0}
     \Big[\big\| v_\phi(\boldsymbol{\theta}_t, t, s)
              - (\tilde{\boldsymbol{\theta}} - \boldsymbol{\theta}_0) \big\|^{2}\Big].
\label{eq:cfm-loss}
\end{equation}

At inference, a base sample $\boldsymbol{\theta}_0$ is drawn and the ODE $\mathrm{d}\boldsymbol{\theta}_t / \mathrm{d}t = v_\phi(\boldsymbol{\theta}_t, t, s)$ is integrated from $t=0$ to $t=1$ with an adaptive Runge--Kutta solver (\texttt{torchdiffeq} \texttt{odeint}, relative and absolute tolerances $10^{-5}$), yielding one posterior sample. We drew 500 Monte-Carlo samples per test light cone.

The CFM network for solving for the velocity field is a stack of seven gated-linear-unit MLP blocks with residual skips, hidden dimension 1024, conditioned at every block on $s$, $t$, and the current $\boldsymbol{\theta}_t$. The 5D parameter targets are linearly scaled to $[0,1]$ using the prior bounds of Table~\ref{tab:target_params}, and the 360D summary is z-score normalized with statistics fit on the training sample only. Each CFM is trained for up to 500 epochs with AdamW (learning rate $10^{-3}$, weight decay $10^{-4}$, batch size 32) and early stopped on the validation loss with patience 8. One CFM fit converges in $1$--$2$\,min and posterior sampling adds another $1$--$2$\,min, so a single (pipeline, budget) cell runs in a few minutes on a single Nvidia A30 GPU and the full sweep is comfortably parallel across ensemble members and budgets.

\subsection{Baseline model comparison}
\label{sec:baselines}

To isolate the effect of the encoder pretraining objective on inference results and transfer performance, all pipelines share the same backbone network architecture. After pretraining, the encoders are frozen and used to summarize each light cone to a 360D vector that is fed into the CFM; only the encoding strategy itself differs:
\begin{itemize}\itemsep0pt
  \item SKATR: the JEPA encoder of Sect.~\ref{sec:skatr}, pretrained once on 67{,}325 noiseless 21cmFAST light cones and frozen for all downstream use.
  \item ViT(21cmFAST): a backbone of the same ViT architecture as SKATR, trained on the same 67{,}325 noiseless 21cmFAST light cones but with a supervised $L_2$ regression objective on the six informative 21cmFAST parameters of Table~\ref{tab:source_params}. This isolates the pretraining objective (self-supervised vs. supervised) from the pretraining data scale. The single backbone ($\sim$20\,h on an Nvidia A100) is reused at every budget $N$.
  \item ViT(Loreli): a backbone of the same ViT architecture, trained supervised end-to-end on the budget-$N$ target light cones with an $L_2$ regression loss on the five Loreli parameters. Since this backbone is trained with Loreli target labels, a new ViT(Loreli) is trained from scratch at every budget $N \in \{100, 500, 1000, 2500, 5000, 6517\}$ ($\sim$3\,h per backbone at $N=6517$ on an Nvidia A100).
\end{itemize}

After each backbone is trained, the regression head is discarded, and the light cone is summarized by mean-pooling the 3072 patch token embeddings into a single 360D vector, identical to the SKATR summarizer. All supervised backbones are trained for 100 epochs with AdamW (learning rate $10^{-3}$, weight decay $10^{-4}$, OneCycle schedule, batch size 64). Each is likewise frozen at its best-validation epoch (Appendix~\ref{app:losses}). For the noise study of Sect.~\ref{sec:results_noise}, the ViT(Loreli) backbone is retrained from scratch on the noisy budget-$N$ Loreli light cones (Sect.~\ref{sec:noise}), giving an in-domain noise-aware supervised baseline; the SKATR and ViT(21cmFAST) encoders remain those pretrained on noiseless 21cmFAST, so their use on noisy data is a genuine zero-shot transfer across the noise boundary. An overview of the three pipelines compared in this work is presented in Fig.~\ref{fig:pipelines}.

\subsection{Ensemble and evaluation protocol}
\label{sec:metrics}

A single CFM training at small label budget $N$ is subject to non-negligible optimization noise from random initialization, where results can vary substantially across seeds. To quantify this effect and average over it, we trained an ensemble of $M\!=\!5$ CFM realizations with independent random initial weights for every (pipeline, budget $N$) cell, sharing the same frozen encoder across the five realizations. Every test light cone therefore yields $M\!=\!5$ posteriors per cell. For each evaluation statistics we report two complementary numbers per cell: the mean $\pm$ standard deviation across the five realizations, which quantifies CFM-related run-to-run scatter, and the combined posterior obtained by concatenating all samples evaluated to Monte-Carlo sample the posterior ($5\!\times\!500 = 2500$ samples per light cone), which is our most accurate point estimate. The combined posterior is what the corner plots and tests of accuracy with random points (TARP) curves in Sect.~\ref{sec:results_clean}--\ref{sec:results_noise} display unless stated otherwise.
We assess three complementary properties on the 1{,}150 held-out test light cones of Sect.~\ref{sec:data}. Accuracy is the per-parameter coefficient of determination $R^2$ between the posterior mean and the truth. We also report the mean of $R^2$ over the set of five inferred parameters. Informativeness is the relative posterior width, defined per parameter as the mean posterior standard deviation across test light cones divided by the standard deviation of the labels (smaller is more informative, with $1$ indicating a posterior as wide as the prior). Calibration is assessed jointly with the TARP coverage test~\citep{Lemos_2023} and marginally with per-parameter rank histograms as a simulation-based calibration diagnostic~\citep[SBC;][]{Talts_2018}. The rank histograms are displayed in Appendix~\ref{app:sbc}.
At each budget the $M$ ensemble members share the same $N$ target light cones, so the error bars in Figs.~\ref{fig:scaling} and~\ref{fig:noisy-scaling} reflect optimization noise alone. Retraining ensembles on independent draws of $N$ light cones (specifically at low $N$) increases the scatter in mean $R^2$, while the scatter in relative posterior width is comparable. The scaling curves are therefore to be read as a single realization of the subsampling. Crucially, the ordering of the pipelines and the label-efficiency gap are preserved across redraws, when pipelines share the same subsample at each budget.

\section{Results I: Cross-simulator transfer}
\label{sec:results_clean}

In this section we discuss testing the noiseless cross-simulator transfer. All three
pipelines reduce a given light cone to a 360D summary,
which is fed into the $M\!=\!5$ CFM ensemble protocol (Sect.~\ref{sec:baselines}). The
encoders are SKATR (self-supervised on $67{,}325$ noiseless 21cmFAST
light cones), ViT(21cmFAST) (supervised on the same source set), and
ViT(Loreli) (supervised on the budget-$N$ target light cones). The encoders
share the same architecture and never see the test split; only the
pretraining objective and the training data scale differ. We address two questions: first, how the five-parameter posteriors and their calibration compare at the full Loreli budget $N=6517$, where every pipeline is at its best; and second, how that comparison evolves as the labeled target data budget $N$ is reduced. Both answers become only more favorable to SKATR under realistic noise in
Sect.~\ref{sec:results_noise}.

\subsection{Posteriors and calibration at full label budget }
\label{sec:posteriors}

\begin{figure*}[h]
    \centering
    \begin{subfigure}{0.63\textwidth}
        \centering
        \includegraphics[width=\linewidth]{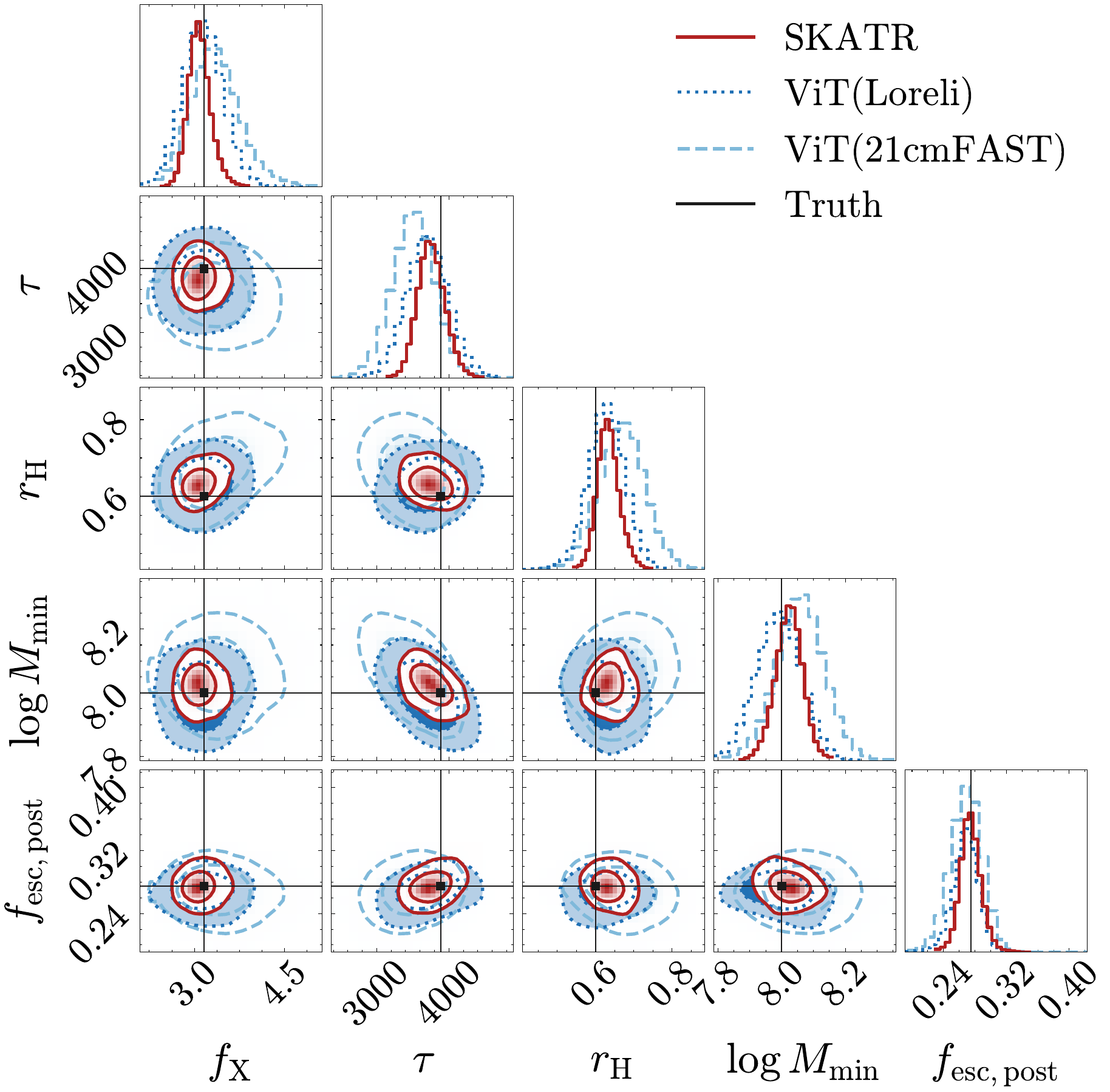}
        \caption{}
        \label{fig:corner-clean}
    \end{subfigure}
    \hfill
    \begin{subfigure}{0.35\textwidth}
        \centering
        \includegraphics[width=\linewidth]{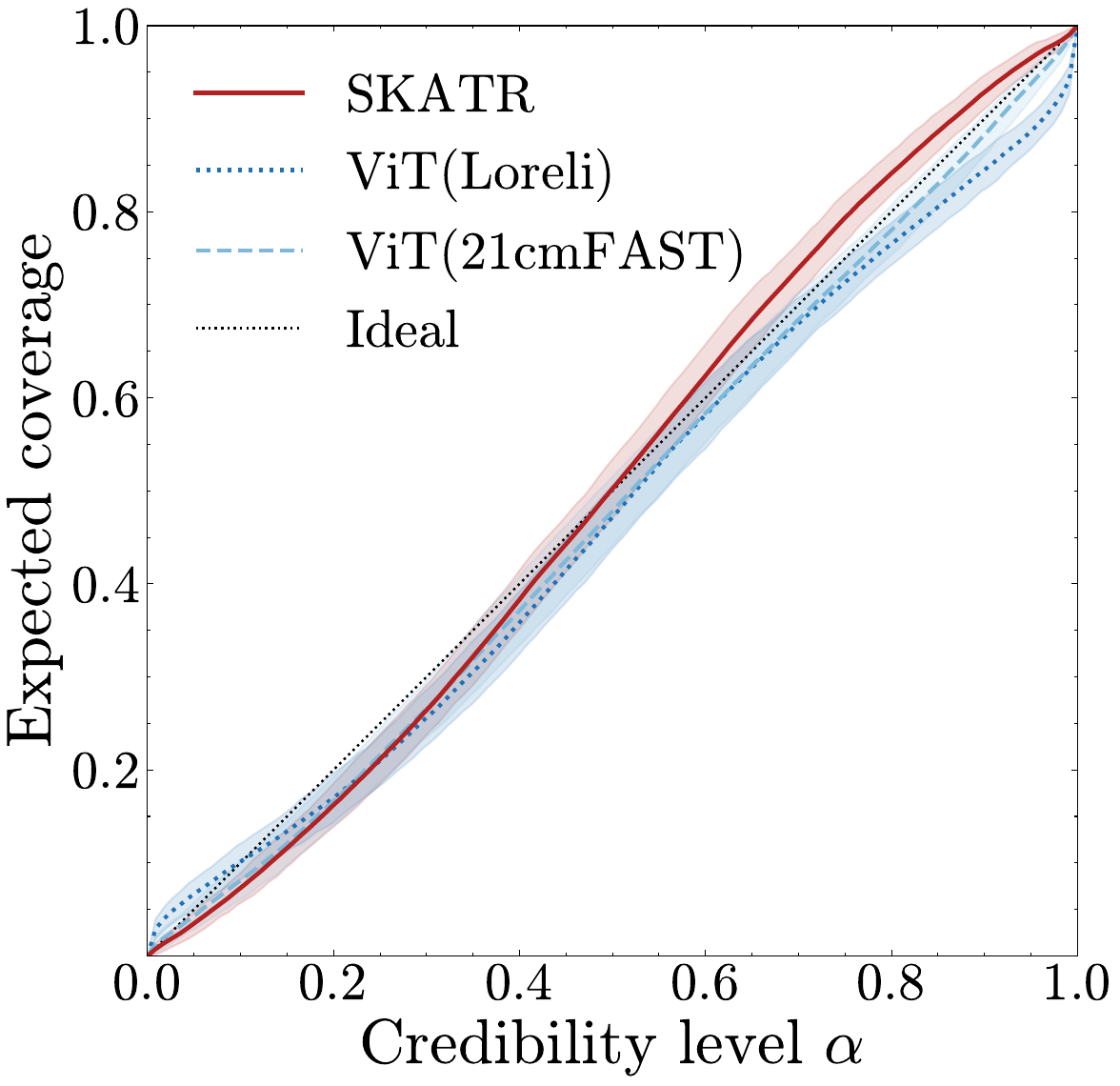}
        \caption{}
        \label{fig:tarp-clean}
    \end{subfigure}
    \caption{Full budget of the labeled Loreli hydrodynamical and RT light cones ($N=6517$), noiseless case. (a) Posterior corner plot for a representative test light cone (truth in black), combined over the ensemble posteriors. The SKATR contours are the tightest and show no signs of bias.
    (b) Joint TARP coverage averaged over the ensemble and test set. SKATR is marginally underconfident, while the two supervised baselines carry a mix of biases, overconfidence, and underconfidence. The per-parameter breakdown becomes clear via the SBC ranks in Appendix~\ref{app:sbc}.}
    \label{fig:posterior-and-calibration}
\end{figure*}

\begin{table*}[t]
\centering
\caption{Per-parameter inference results.}
\label{tab:uncertainty_r2}
\small
\setlength{\tabcolsep}{3.5pt}
\begin{tabular}{lcccccc}
\toprule
& \multicolumn{3}{c}{$R^2$ ($\uparrow$)} & \multicolumn{3}{c}{Rel.\ unc.\ ($\downarrow$)} \\
\cmidrule(lr){2-4}\cmidrule(lr){5-7}
Parameter            & ViT(Loreli) & ViT(21cmFAST) & SKATR          & ViT(Loreli) & ViT(21cmFAST) & SKATR \\
\midrule
$f_X$                 & 0.97\,(0.019) & 0.97\,(0.003) & \textbf{0.99}\,(0.002) & 0.12\,(0.018) & 0.11\,(0.015) & \textbf{0.06}\,(0.008) \\
$\tau$                & 0.95\,(0.006) & 0.94\,(0.006) & \textbf{0.99}\,(0.002) & 0.14\,(0.024) & 0.18\,(0.022) & \textbf{0.08}\,(0.013) \\
$r_\mathrm{H}$        & 0.92\,(0.004) & 0.88\,(0.006) & \textbf{0.97}\,(0.004) & \textbf{0.14}\,(0.024) & 0.25\,(0.021) & \textbf{0.14}\,(0.015) \\
$\log M_\mathrm{min}$ & 0.98\,(0.003) & 0.98\,(0.003) & \textbf{1.00}\,(0.001) & 0.11\,(0.014) & 0.13\,(0.015) & \textbf{0.06}\,(0.008) \\
$f_\mathrm{esc,post}$ & 0.95\,(0.005) & 0.91\,(0.012) & \textbf{0.97}\,(0.003) & 0.12\,(0.032) & 0.20\,(0.029) & \textbf{0.09}\,(0.017) \\
\midrule
mean                  & 0.96\,(0.007) & 0.94\,(0.005) & \textbf{0.98}\,(0.001) & 0.12\,(0.022) & 0.17\,(0.020) & \textbf{0.08}\,(0.012) \\
\bottomrule
\end{tabular}
\tablefoot{
Metrics are the coefficient of determination $R^2$ ($\uparrow$ higher is better) and relative posterior width (defined as the mean posterior standard deviation divided by the standard deviation of the label priors; $\downarrow$ lower is more informative) at the full Loreli label budget ($N=6517$). Values are computed on the combined ensemble posterior; the number in brackets is the standard deviation across the $M=5$ CFM realizations (Sect.~\ref{sec:metrics}), which measures run-to-run optimization noise. The best value per row is marked in bold.
}
\end{table*}

At the full budget of 6517 labeled Loreli~II light cones for transfer, all three pipelines recover the truth but with markedly different posteriors. Figure~\ref{fig:posterior-and-calibration}(a) shows the combined-ensemble posterior (see Sect.~\ref{sec:metrics}) for a representative test light cone: the SKATR contours are the tightest and stay centered on the truth, while the two supervised baselines are visibly broader. The pattern is systematic across the test set (Table~\ref{tab:uncertainty_r2}): SKATR attains the best $R^2$ on each of the five parameters, with a mean of $0.98$, and the smallest relative posterior width (posterior standard deviation divided by the standard deviation of the label priors) on all five, tying ViT(Loreli) at $0.14$ on $r_\mathrm{H}$ and strictly narrowest on the other four. The advantage is largest on the two parameters the supervised baselines find most difficult to constrain, which are the hard X-ray fraction $r_\mathrm{H}$ and the late-time escape fraction $f_\mathrm{esc,post}$. Notably, these are the ones that are also the most sparsely sampled parameters within the Loreli~II database (see Sect.~\ref{sec:loreli}). These gaps are large compared with the run-to-run scatter of the $M=5$ CFM realizations, quoted in brackets in Table~\ref{tab:uncertainty_r2}. Measured in units of that scatter, SKATR exceeds ViT(Loreli) in $R^2$ by three to ten standard deviations on $\tau$, $r_\mathrm{H}$, $\log M_\mathrm{min}$ and $f_\mathrm{esc,post}$, and by $1.4$ standard deviations on $f_X$; the differences in posterior width are less pronounced, between $0.2$ and $2.5$ standard deviations, with ViT(Loreli) and SKATR statistically indistinguishable on $r_\mathrm{H}$.

Two points deserve further emphasis. First, the ViT(21cmFAST) pipeline is already a strong inference baseline for transfer from training on 21cmFAST semi-numerical simulations to the Loreli database of hydrodynamical and RT light cones, reaching a mean $R^2$ of $0.94$: meaning a supervised encoder pretrained on a large source dataset with a suitable 3D ViT architecture and a sufficiently large (360D) embedding (data summary) transfers to the target simulator without ever seeing a Loreli label. To a certain extent, cross-simulator transfer is therefore not restricted to the self-supervised objective, although differences in precision and posterior calibration persist, stemming from the quality of the data embeddings. Second, and against that already well-constraining baseline, the JEPA self-supervised pretraining adds substantial further gains: from $0.94$ to $0.98$ in mean $R^2$ pointing to unbiased constraints across the prior range, with the largest absolute gains on $r_\mathrm{H}$ and $f_\mathrm{esc,post}$ parameters, together with uniformly more precise posteriors. The self-supervised pretraining objective, not merely the
pretraining data scale, is what results in a more optimal data embedding (summary) that enables both unbiased and more precise constraints.

Going beyond an assessment of precision and bias for the transferred parameter estimates, we assess calibration of the full posteriors with the joint TARP coverage test~\citep{Lemos_2023} (Fig.~\ref{fig:posterior-and-calibration}b) and marginal calibration with per-parameter SBC rank histograms (Appendix~\ref{app:sbc}).
SKATR shows a small indication of underconfidence, an observations already made in~\citet{Ore_2025}. The supervised baselines show signs of bias (both pipelines) and overconfidence (ViT(Loreli)). The SBC ranks (Appendix Fig.~\ref{fig:sbc-clean}) resolve the picture per parameter: SKATR and ViT(21cmFAST) are flat (calibrated) on four of the five parameters and mildly $\cap$-shaped (underconfident, i.e.,\ posteriors slightly wider than the residuals require) on $f_\mathrm{esc,post}$. ViT(Loreli) shares the $f_\mathrm{esc,post}$ underconfidence but is visibly $\cup$-shaped on $r_\mathrm{H}$ (posteriors too narrow, i.e.,\ overconfident). Among the three pipelines, only ViT(Loreli) shows significant overconfidence on the $r_\mathrm{H}$ parameter; SKATR is perfectly calibrated on four parameters and slightly conservative on one.

The shared underconfidence on $f_\mathrm{esc,post}$ likely has a common origin in the sparseness of its labels and continuous NPE: the CFM transports a Gaussian base into the posterior via a smooth continuous-time ODE, which makes it hard to concentrate mass tightly on the three isolated values that $f_\mathrm{esc,post}$ takes in the Loreli~II database (Table~\ref{tab:target_params}). All three pipelines therefore return posteriors somewhat wider than the discrete-label structure warrants, which is an underconfident rather than overconfident deviation and hence the safer failure mode for miscalibration. SKATR's calibration advantage lies in the other four parameters, and as we show in Sect.~\ref{sec:results_noise} this advantage sharpens further under realistic noise.
In summary, the supervised pipelines tend to be overconfident (meaning their posteriors should be larger) and biased, while at the same time their precision and $R^2$ values are reduced relative to SKATR. This points to a less-than-optimal embedding, while our self-supervised training objective and thus embedding found with SKATR enables more precise constraints (higher information content encoded) and calibrates better (no overconfidence or biases detected).

\subsection{Data scaling and simulation budget efficiency}
\label{sec:scaling}
\begin{figure*}[h]
\sidecaption
    \centering
    \includegraphics[width=12cm]{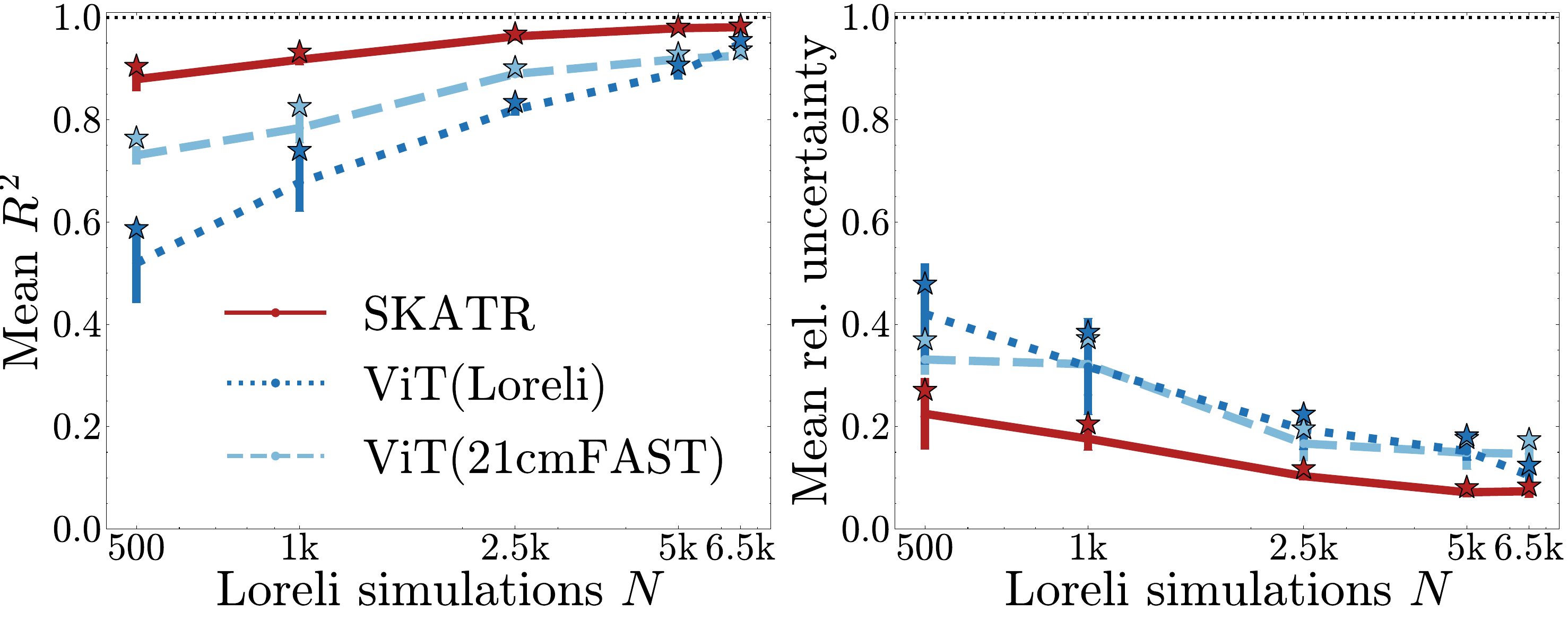}
    \caption{Scaling with the number of Loreli training simulations $N$: accuracy (left, mean $R^2$) and posterior informativeness (right, mean relative uncertainty; lower is more informative). The error bars show the spread over the five-member CFM ensembles. The stars mark the combined ensemble. SKATR leads at every budget $N$ of the Loreli light cones shown and exceeds the full-budget supervised (trained on $\sim$6.5k simulations) baseline at $N=2.5$k.}
    \label{fig:scaling}
\end{figure*}

Figure~\ref{fig:scaling} shows how the three pipelines behave as the Loreli training budget $N$ of simulations available is reduced from $6517$ down to $500$ in five steps. Both encoders pretrained on the large low-cost source data (SKATR and ViT(21cmFAST)) substantially outperform the direct ViT(Loreli) fit, which trains its encoder from scratch on the limited target Loreli data and overfits most severely where labels are scarcest. At $N=500$ the combined-ensemble accuracies are $R^2=0.90$ (SKATR) and $0.76$ (ViT(21cmFAST)), both well above $0.59$ for ViT(Loreli). ViT(Loreli) only reaches the accuracy of ViT(21cmFAST) around the full budget of target training simulations, reaching $R^2=0.96$ at $N=6517$ while still falling short of SKATR's $0.98$. SKATR stays ahead of ViT(21cmFAST) at every training budget tested ($0.90$ and $0.76$ at $N=500$, $0.93$ and $0.83$ at $N=1000$, narrowing to $0.98$ and $0.94$ at full budget, respectively) and shows the smallest run-to-run scatter across the $M=5$ random initializations for ensembling at every budget. The calibration ordering of Sect.~\ref{sec:posteriors} persists across all budgets: SKATR's joint TARP stays calibrated to mildly conservative at every simulation budget.

In practice, this translates into significant savings in the number of expensive target simulations. SKATR trained on $N=2500$ Loreli labeled simulations reaches $R^2=0.97$, already ahead of the fully supervised ViT(Loreli) trained on the entire pool of $6517$ labels ($R^2=0.96$): a $2.6\times$ reduction in target-simulator labels for equal or better accuracy. In forward-modeling cost (Sect.~\ref{sec:data}) this is $\sim\!1.6\times10^6$\,CPU-hours of Loreli~II hydrodynamical simulations avoided, against a one-time $\sim\!1.7\times10^4$\,CPU-h cost for the 21cmFAST pretraining set and $\sim$110\,GPU-h for SKATR pretraining that are amortized across every downstream task budget. The supervised ViT(Loreli) accuracy gap  can be closed further with target-specific measures such as augmentation, regularization, or hyperparameter tuning. In Appendix~\ref{app:aug} we study rotate-and-reflect (R+R) augmentations on the ViT(Loreli) backbone, a physically natural choice for 21cm light cones, and find that the augmented baseline reaches accuracy comparable to SKATR at the full budget. In the label-scarce regime, however, SKATR retains a clear data-efficiency edge that target-side augmentation does not close (Fig.~\ref{fig:scaling-appendix}). The key distinction of SKATR is independent of where such tuning ends up: it reaches the same accuracy using a single frozen encoder, pretrained on approximate 21cmFAST light cones and never exposed to the target simulator or the noise model, which demonstrates superior generalization. This parameterization- and noise-agnostic property is what carries into the realistic mock regime of Sect.~\ref{sec:results_noise}.

\section{Results II: Cross-simulator transfer under realistic SKA noise}
\label{sec:results_noise}

After the first domain shift, transferring from semi-numerical to hydrodynamical and RT simulations, we added a second one to the target Loreli light cones: the SKA AA$^\ast$ thermal noise and foreground-wedge cut described in Sect.~\ref{sec:noise}. For SKATR and ViT(21cmFAST) this is a strict zero-shot transfer: both encoders were pretrained to summarize noiseless light cones simulated with 21cmFAST and were subsequently applied as frozen models to summarize noisy hydrodynamical data, having seen neither the target physics of Loreli nor the noise. As the strongest supervised competitor to benchmark against, we added the ViT(Loreli) model of Sect.~\ref{sec:baselines} but as a fresh backbone retrained from scratch on the noisy budget-$N$ Loreli light cones and dropped the noiseless ViT(Loreli) baseline. We addressed the same two questions as in Sect.~\ref{sec:results_clean}: how the pipelines, both fully supervised but trained on fewer Loreli light cones and SKATR with zero-shot transfer, compare at the full Loreli budget, and how that comparison evolves as the number of Loreli light cones available for training shrinks.

\subsection{Posteriors and calibration at the full budget}
\label{sec:noisy-posteriors}

\begin{figure*}[h]
    \centering
    \begin{subfigure}{0.63\textwidth}
        \centering
        \includegraphics[width=\linewidth]{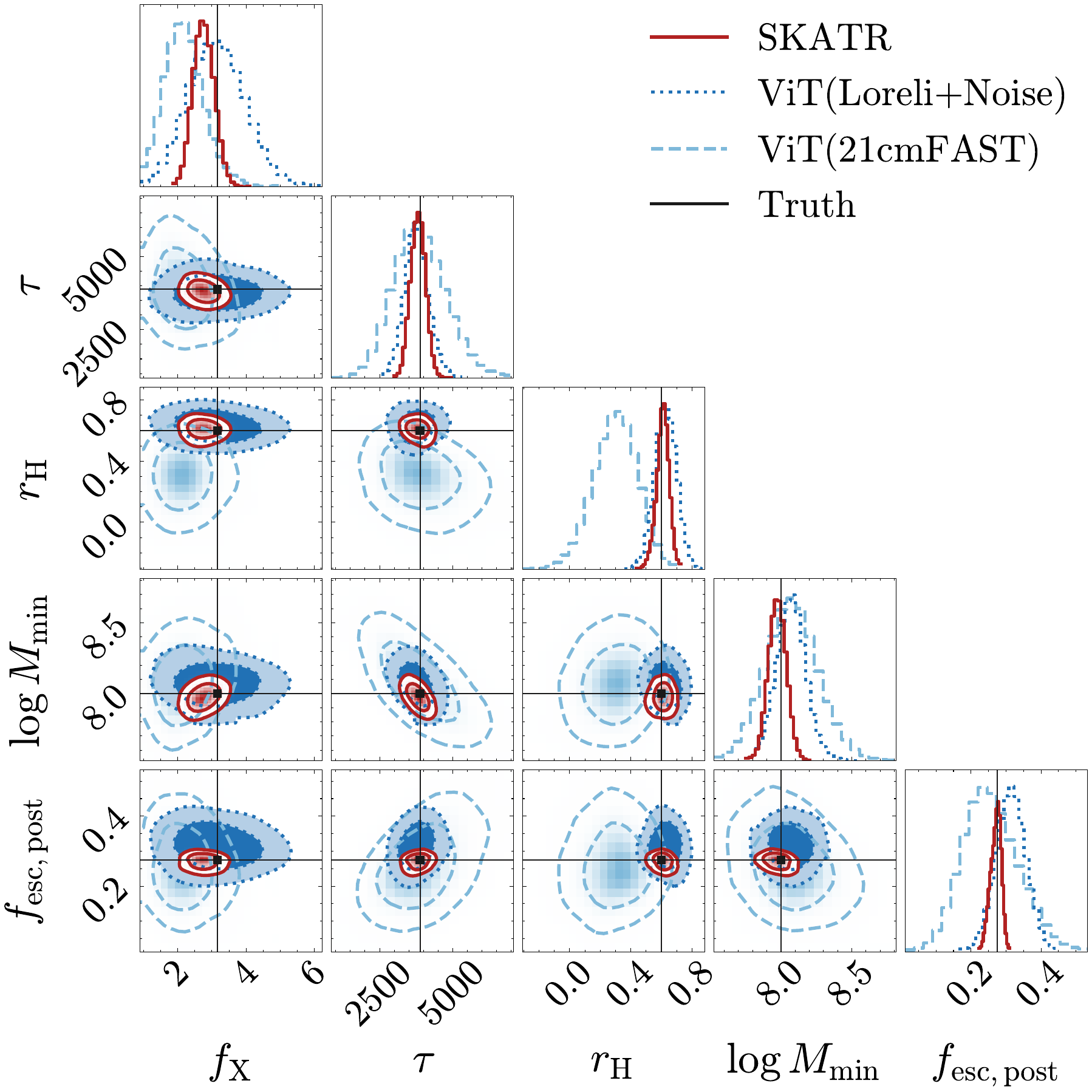}
        \caption{}
        \label{fig:corner-noisy}
    \end{subfigure}
    \hfill
    \begin{subfigure}{0.35\textwidth}
        \centering
        \includegraphics[width=\linewidth]{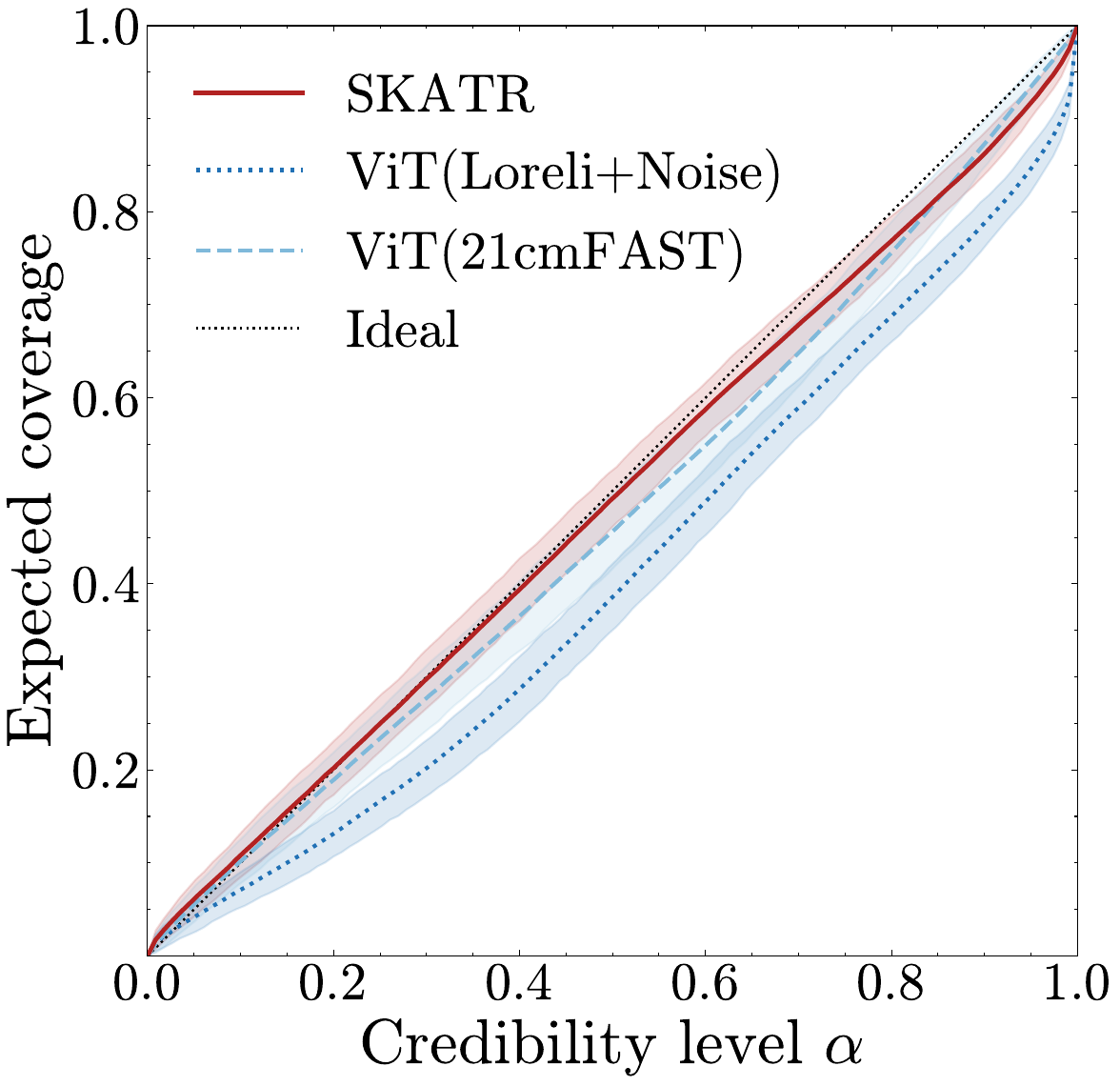}
        \caption{}
        \label{fig:tarp-noisy}
    \end{subfigure}
    \caption{Inference under the full Loreli budget of training light cones ($N=6517$) with SKA AA$^\ast$ noise. (a) Posterior corner plot for a representative test light cone (truth in black), combined over the $M=5$ CFM ensemble. SKATR is the tightest, showing no signs of bias. The in-domain noise-trained ViT(Loreli) shows slightly broader constraints, and the zero-shot ViT(21cmFAST) is broad and sits off the truth on several parameters. (b) Joint TARP coverage averaged over the ensemble and test set. SKATR tracks the diagonal (well-calibrated) and the noise-trained ViT(Loreli) bows below it (overconfident and bias). The zero-shot ViT(21cmFAST) lies near the diagonal because its posteriors are so broad and uninformative that the posteriors trivially achieve good coverage.}
    \label{fig:posterior-and-calibration-noisy}
\end{figure*}

\begin{table*}[t]
\centering
\caption{Per-parameter inference results under SKA AA$^\ast$ noise.}
\label{tab:r2_noisy}
\small
\setlength{\tabcolsep}{3.5pt}
\begin{tabular}{lcccccc}
\toprule
& \multicolumn{3}{c}{$R^2$ ($\uparrow$)} & \multicolumn{3}{c}{Rel.\ unc.\ ($\downarrow$)} \\
\cmidrule(lr){2-4}\cmidrule(lr){5-7}
Parameter            & ViT(Loreli, noise-trained) & ViT(21cmFAST) & SKATR          & ViT(Loreli, noise-trained) & ViT(21cmFAST) & SKATR \\
\midrule
$f_X$                 & 0.94\,(0.065) & 0.84\,(0.034) & \textbf{0.96}\,(0.003) & 0.26\,(0.031) & 0.26\,(0.020) & \textbf{0.11}\,(0.013) \\
$\tau$                & 0.87\,(0.009) & 0.66\,(0.024) & \textbf{0.92}\,(0.008) & 0.21\,(0.026) & 0.44\,(0.023) & \textbf{0.19}\,(0.027) \\
$r_\mathrm{H}$        & 0.89\,(0.011) & 0.70\,(0.018) & \textbf{0.92}\,(0.004) & 0.20\,(0.027) & 0.44\,(0.027) & \textbf{0.19}\,(0.022) \\
$\log M_\mathrm{min}$ & 0.94\,(0.005) & 0.80\,(0.011) & \textbf{0.97}\,(0.004) & 0.19\,(0.028) & 0.37\,(0.013) & \textbf{0.14}\,(0.016) \\
$f_\mathrm{esc,post}$ & 0.87\,(0.008) & 0.69\,(0.008) & \textbf{0.91}\,(0.008) & 0.22\,(0.028) & 0.46\,(0.031) & \textbf{0.20}\,(0.030) \\
\midrule
mean                  & 0.91\,(0.008) & 0.74\,(0.017) & \textbf{0.94}\,(0.004) & 0.22\,(0.025) & 0.40\,(0.020) & \textbf{0.17}\,(0.022) \\
\bottomrule
\end{tabular}
\tablefoot{Metrics are the coefficient of determination $R^2$ ($\uparrow$) and relative posterior width ($\downarrow$) at the full Loreli budget of training light cones ($N=6517$) under SKA AA$^\ast$ noise. Values are from the combined ensemble posterior; the brackets give the standard deviation across the $M=5$ CFM realizations (Sect.~\ref{sec:metrics}). The best value per row is in bold.}
\end{table*}

Under noise the three pipelines fall into three distinct regimes, visible in the representative posteriors of Fig.~\ref{fig:posterior-and-calibration-noisy}(a) and across the test set in Table~\ref{tab:r2_noisy}. First, zero-shot supervised transfer is fragile. The frozen ViT(21cmFAST) encoder, applied to noisy Loreli data without ever having seen Loreli light cones or noise, drops to mean $R^2 = 0.74$ with relative posterior widths around $0.40$. The supervised representation appears to have learned small-scale source-specific features that are washed out by the noise and the wedge cut. Notably, the constraints on $f_X$ are the least degraded, indicating that the X-ray heating signal might be more robust to the noise simulator-specific small-scale behavior than the other parameters.

Secondly, the in-domain ViT trained on noisy Loreli light cones maintains accuracy. Retraining the ViT(Loreli) encoder on the noisy budget-$N$ data yields mean $R^2$ of $0.91$ at the full budget, only $0.03$ short of SKATR's $0.94$, with relative posterior widths around $0.22$. But this recovery comes at a price: the joint TARP curve of Fig.~\ref{fig:posterior-and-calibration-noisy}(b) drops below the diagonal, indicating a bias, and the per-parameter rank histograms (SBC; Appendix~\ref{app:sbc}, Fig.~\ref{fig:sbc-noisy}) are $\cup$-shaped on several parameters, indicating that the noise-trained ViT(Loreli) posteriors are systematically too narrow, making it overconfident.

Finally, SKATR is accurate, informative, and calibrated. The frozen SKATR encoder, blind to both the simulator and the noise, reaches mean $R^2 = 0.94$ with the smallest relative posterior widths on every parameter (mean $0.17$), and its joint TARP lies close to the diagonal. The per-parameter SBC ranks are essentially flat with a small indication of overconfidence for $r_\mathrm{H}$. The slight underconfidence for $f_\mathrm{esc,post}$ mentioned in Sect.~\ref{sec:posteriors} is more reduced. SKATR is therefore the only pipeline in this comparison that is simultaneously accurate, precise, and calibrated, and it achieves all three without the encoder ever having seen the simulator or the noise model.

\subsection{Data scaling and robustness}
\label{sec:noisy-scaling}

\begin{figure*}[h]
    \centering
    \sidecaption
    \includegraphics[width=12cm]{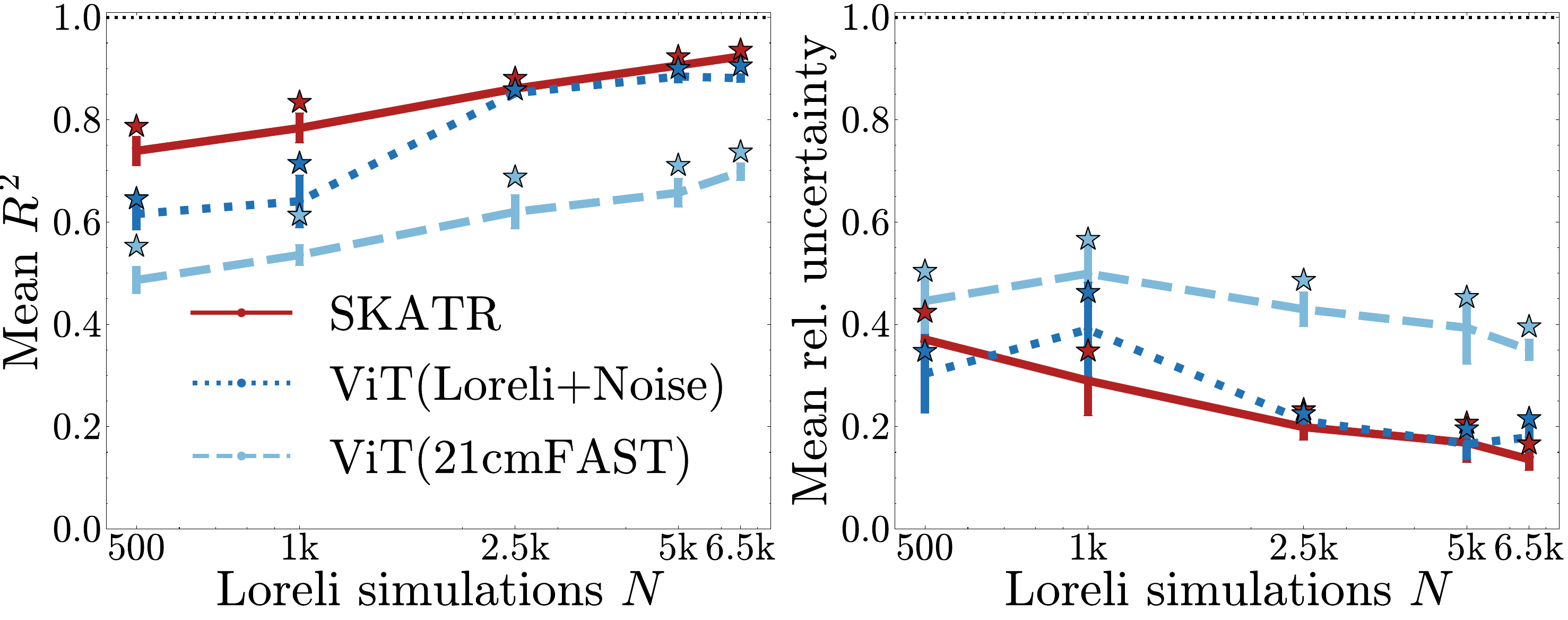}
    \caption{Scaling with the number of Loreli training simulations $N$ under SKA AA$^\ast$ noise: accuracy
             (left, mean $R^2$) and posterior informativeness (right, mean
             relative uncertainty; lower is more informative). The error bars show
             the spread over the five-member CFM ensembles. The stars mark the
             combined ensemble. SKATR leads at every budget.}
    \label{fig:noisy-scaling}
\end{figure*}

Figure~\ref{fig:noisy-scaling} shows the pipeline performance, as the training budget $N$ of Loreli light cones is reduced from $N=6517$ to $N=500$ under noise. SKATR again leads at every budget, rising from mean $R^2 = 0.79$ at $N=500$ to $0.94$ at the full budget; the in-domain noise-trained ViT(Loreli) follows ($0.65 \to 0.91$) and the zero-shot ViT(21cmFAST) trails ($0.55 \to 0.74$). The largest gaps between SKATR and the noise-trained baseline are in the label-scarce regime ($0.79$ vs $0.65$ at $N=500$, $0.83$ vs $0.71$ at $N=1000$) closing in as the noise-trained supervised ViT(Loreli) absorbs more target labels.

The impact of the pretraining objective is isolated most cleanly by the two pipelines that share the same architecture, pretraining data, and downstream protocol, and that differ only in the self-supervised versus supervised pretraining objective: SKATR and ViT(21cmFAST). With encoders fixed and only the target input distribution changed from noiseless to noisy Loreli light cones, the full-budget mean $R^2$ degrades as $0.98 \to 0.94$ for SKATR (a $4\%$ drop) and as $0.94 \to 0.74$ for ViT(21cmFAST) (a $21\%$ drop). SKATR's JEPA features are therefore close to five times more robust to the same input shift than the supervised features of ViT(21cmFAST), pinning the pretraining objective itself as the source of the noise robustness rather than the data scale, the architecture, or the downstream protocol.

\section{Discussion and conclusions}
\label{sec:conclusion}

\paragraph{Summary:} This work establishes SKATR as a foundation model for reionization inference: pretrained once, label-free, on low-cost noiseless 21cmFAST light cones, then frozen, it transfers under a joint domain shift to a different, hydrodynamical simulator that explicitly resolves RT and to realistic SKA~AA$^\ast$ thermal noise with foreground-wedge removal. It has seen neither the target simulator, nor its parameters, nor the noise model. On noiseless Loreli~II data, SKATR reaches a mean $R^2 = 0.98$ against $0.96$ for the supervised ViT(Loreli) at the same budget of training simulations, with the tightest posterior on all five parameters. Under realistic SKA~AA$^\ast$ thermal noise plus wedge avoidance, the same frozen encoder retains $96\%$ of its noiseless accuracy (mean $R^2$ drops from $0.98$ to $0.94$) and exceeds the in-domain noise-trained ViT(Loreli) on every parameter, despite never having seen the target simulator, its parameter set, or the noise realization. In both regimes SKATR additionally remains well calibrated, while the supervised pipelines show signs of miscalibration, making it the only pipeline in our comparison that is simultaneously accurate, informative, and calibrated. Beyond accuracy, the foundation-model approach amortizes pretraining across all downstream tasks: at $N=2500$ target light cones, SKATR already matches the fully supervised ViT(Loreli) at $N=6517$, saving a factor of $2.6$ in target simulations and $\sim\!1.6\times10^6$\,CPU-h of RT hydrodynamical simulations, for a one-time investment of $\sim\!1.7\times10^4$\,CPU-h (source set) and $\sim\!110$\,GPU-h (pretraining). The instrument and the reionization physics behind any near-term SKA measurement will both differ from the training simulator, and a frozen, label-free summarizer that remains calibrated under such unseen shifts is precisely the property one wants before applying SBI to data. It achieves this with a label-free, symmetry-agnostic objective: unlike the augmented baseline of Appendix~\ref{app:aug}, no symmetry group is chosen for it in advance.

\paragraph{Why JEPA transfers:} We can only speculate about why JEPA features transfer across both the simulator and the noise shift before opening up and diagnosing the latent space itself, but the two robustness results plausibly share a common mechanism. Predicting masked blocks of voxels in embedding space, for families of both smaller- and larger-scale masks, rewards an encoder that captures correlations across scales together with the connections between them. It therefore encourages a representation organized around the physics of the signal rather than around any single scale. Supervised $L_2$ regression, by contrast, is free to collapse features onto whatever simulator-specific structures minimize the parameter-regression loss on the training set. This training set must be neither physically meaningful nor shared across simulators. A representation tied to the underlying physics should then retain much of its information when the input is perturbed -- whether by a change of simulator, by incoherent thermal noise, or by the removal of Fourier modes inside the foreground wedge. By contrast, one tuned to incidental discriminative features can lose it completely. This would explain why the same encoder transfers across simulators and remains robust to noise without either having been a training target. Predicting in embedding space rather than in voxel space is what we expect to provide transferability. Voxel reconstruction can force the encoder to retain small-scale details, including the simulator-specific texture and noise-like modes, whereas an embedding-space target can discard what is irrelevant in terms of physical predictions. It instead retains structure across scales in relation to physical predictions. This is consistent with the controlled comparisons discussed in Sect.~\ref{sec:skatr}, and with the analysis of \citet{Littwin_2024}, who show that latent-space prediction carries an implicit bias toward high-influence features, away from noisy ones. We stress that this is a working hypothesis: confirming it requires a direct latent-space analysis of what SKATR has actually learned, which we identify below as the most important direction for future work.

\paragraph{Implications for inference from the cosmological large-scale structure:} The decisive property for SKA-era large-scale structure inference is not the mechanism behind SKATR but its consequence: SKATR's pretraining does not know anything about the target simulator or the noise model, and it does not have to. Real observations in general and real SKA AA$^\ast$ observations in particular will differ from any training simulator in both physics and systematics. No supervised pipeline can be perfectly tuned in advance to a domain that has not been observed. As our results in Sect.~\ref{sec:results_noise} demonstrate, a single frozen encoder applied zero-shot across a joint simulator-and-noise shift achieves better accuracy, informativeness, and calibration than a baseline whose encoder was retrained on the noisy target. To our knowledge this is the first demonstration of a cross-simulator and noise transfer for a cosmological large-scale structure signal, and it provides a realistic, trustable path forward for applying SBI to actual SKA data once they arrive.

\paragraph{Position in the landscape:} Some recent studies have demonstrated foundation-model-style self-supervised pretraining for astronomy and cosmology with a focus on multimodal alignment for galaxy surveys \citep[e.g.,][]{Smith_2024, Parker_2024, Parker_2025}. However, robust cosmological inference from large-scale structures, in particular the 21cm tomographic signal, has so far been approached either by supervised neural summaries linked to one simulator \citep{Prelogovic_2024, Schosser_2026} or by power-spectrum analyses that discard the non-Gaussian information~\citep{Saxena_2023, Greig_2024, Pietschke_2025, Cooper_2026}. Domain adaptation provides an alternative for known-target settings \citep{Roncoli_2023, Ntampaka_2025, Pierre_2026}. However, it requires labeled or paired target data that we will not have for SKA, at least initially, and will remain expensive to generate. Our compute-asymmetric recipe, which uses a low-cost semi-numerical pretraining set (here 21cmFAST, though equally applicable to other models), generalizes immediately to any forthcoming SKA-targeted high-fidelity simulation suite, such as the hydrodynamical and RT Loreli~II database and \texttt{pyC2Ray}~\citep{Hirling_2023}. Each of these target simulators offers a different combination of astrophysical parameterization, resolution, and parameter coverage; the same frozen SKATR encoder should remain a valid summary for all of them.

\paragraph{Scope and outlook:} We note a few limitations of and possible next steps for this work. First, our noise model captures thermal noise from the SKA~AA$^\ast$ array configuration plus a horizon-buffer wedge cut but not the full calibration chain, residual foreground subtraction errors, or beam systematics. Extending the demonstrated agnosticism to these systematics is a natural next step. Since SKATR demonstrates transfer from noiseless to noisy data, we expect this to be within scope. In addition, we used a single target simulator family (Loreli~II). A comparative study for testing robustness to further simulator families is implied by our self-supervised pretraining mechanism but remains untested. Furthermore, our comparison is a controlled ablation in which all encoders share the same architecture and training protocol, so that differences are attributable to the pretraining objective rather than to per-pipeline tuning. In Appendix~\ref{app:aug} we relax this constraint for the supervised baselines and find that the ViT(Loreli) can reach SKATR-level accuracy at the full budget with additional R+R augmentation, at the cost of choosing the augmentation group with the target-domain knowledge, which SKATR does not require. In the label-scarce regime, SKATR's data-efficiency advantage persists even against the augmented baseline. The encoder size was likewise held fixed throughout. Depth, number of attention heads, and $360$D embedding were inherited unchanged from \citet{Ore_2025}, where this configuration was selected, and are shared by all three pipelines so that the pretraining objective remains the only variable. Scanning the encoder parameter count would naturally complement the data-scaling results in Figs.~\ref{fig:scaling} and~\ref{fig:noisy-scaling}. In this architecture, however, the embedding width and dimension of the data summary are the same quantity, since the summary is the mean of the token embeddings (Sect.~\ref{sec:skatr}). Changing the model size would simultaneously change the dimensionality of the vector handed to the inference head. Such a scan would therefore confound encoder capacity with summary capacity instead of isolating one from the other. A more elaborate experiment would involve separating the two for a projection, which holds the summary dimension fixed while the encoder grows, or for sweeping the two independently. We leave this to future work, together with the related question of how far the summary can be compressed before transfer degrades.
We also note that instead of a regression pretraining task, one could pretrain the encoder jointly with the CFM for an end-to-end posterior estimation. Since this approach would directly capture uncertainties during pretraining, we expect this to somewhat improve our supervised baseline pipelines, both in-domain (ViT(Loreli)) and out-of-domain ViT(21cmFAST). Similarly, the pretrained SKATR encoder could itself be fine-tuned jointly with the CFM task head, so the scope for such tuning is not restricted to the supervised pipelines. We defer this comparison to future work.

The most interesting follow-up study, however, is a
direct latent-space analysis: opening up the 360D SKATR representation to for example ask whether its summaries actually cluster by physical reionization regime
(e.g.,\ by $\bar{x}_{\mathrm{HI}}(z)$, by the timing of bubble overlap,
by IGM thermal history) rather than by simulator family or noise artifact. Such an analysis would directly test the morphology (physics) versus texture (simulator) hypothesis above and pin down the features that make the transfer work, providing not only a deeper understanding of SKATR's underlying mechanism but also concrete handles for sharpening reionization inference more broadly.
Beyond that, joint cross-probe SBI in which the SKATR summary is one input alongside CMB or high-redshift galaxy observables \citep[in the spirit of][]{Schosser_2026, Pietschke_2026} to anchor high-redshift physics
is another path for future work. In conclusion, self-supervised pretraining on
fast semi-numerical simulations is a viable route to calibrated,
simulator- and noise-agnostic SBI for SKA-era reionization studies as well as cosmological large-scale structure studies more broadly, and
the recipe demonstrated here is ready to be deployed on real SKA data
as the array assembly approaches completion.

\section*{Data availability}
This work builds upon the SKA Transformer (SKATR). The code related to the transfer extension is publicly available at  \href{https://github.com/ypietschke/skatr-transfer}{github.com/ypietschke/skatr-transfer}. The trained encoder weights, together with the training configurations, are archived on Zenodo at \href{https://doi.org/10.5281/zenodo.23597896}{doi.org/10.5281/zenodo.23597896}.
Inference was performed with the \texttt{EoRFlow} framework, available at \href{https://github.com/astro-ML/EoRFlow}{github.com/astro-ML/EoRFlow}.

\begin{acknowledgements}
We thank Divesh Jain, Vrund Patel and M. Paola Vaccaro for useful discussions. We also thank the anonymous referee for their insightful comments, which improved the clarity and robustness of this work. YP's and CH's work is funded by the Volkswagen Foundation and supported through the Federal Ministry of Education and Research (BMBF) and the Ministry of Science, Research and the Arts of Baden-Württemberg. CH also acknowledges funding via the SciFM consortium (05D25VH3) funded by the German Federal Ministry of Research, Technology, and Space (BMFTR) in the ErUM-Data action plan. This work was supported by the DFG under Germany's Excellence Strategy EXC 2181/1 - 390900948 The Heidelberg STRUCTURES Excellence Cluster. The authors acknowledge support by the High Performance and Cloud Computing Group at the Zentrum f\"ur Datenverarbeitung of the University of T\"ubingen, the state of Baden-W\"urttemberg through bwHPC and the German Research Foundation (DFG) through grant no INST 37/935-1 FUGG.
\end{acknowledgements}

\bibliographystyle{aa}
\bibliography{references}

@article{Parsons_2010,
doi = {10.1088/0004-6256/139/4/1468},
url = {https://dx.doi.org/10.1088/0004-6256/139/4/1468},
year = {2010},
month = {mar},
publisher = {The American Astronomical Society},
volume = {139},
number = {4},
pages = {1468},
author = {Aaron R. Parsons and Donald C. Backer and Griffin S. Foster and Melvyn C. H. Wright and Richard F. Bradley and Nicole E. Gugliucci and Chaitali R. Parashare and Erin E. Benoit and James E. Aguirre and Daniel C. Jacobs and Chris L. Carilli and David Herne and Mervyn J. Lynch and Jason R. Manley and Daniel J. Werthimer},
title = {THE PRECISION ARRAY FOR PROBING THE EPOCH OF RE-IONIZATION: EIGHT STATION RESULTS},
journal = {\aj}
}

@ARTICLE{Tingay_2013,
       author = {{Tingay}, S.~J. and {Goeke}, R. and {Bowman}, J.~D. and {Emrich}, D. and {Ord}, S.~M. and {Mitchell}, D.~A. and {Morales}, M.~F. and {Booler}, T. and {Crosse}, B. and {Wayth}, R.~B. and {Lonsdale}, C.~J. and {Tremblay}, S. and {Pallot}, D. and {Colegate}, T. and {Wicenec}, A. and {Kudryavtseva}, N. and {Arcus}, W. and {Barnes}, D. and {Bernardi}, G. and {Briggs}, F. and {Burns}, S. and {Bunton}, J.~D. and {Cappallo}, R.~J. and {Corey}, B.~E. and {Deshpande}, A. and {Desouza}, L. and {Gaensler}, B.~M. and {Greenhill}, L.~J. and {Hall}, P.~J. and {Hazelton}, B.~J. and {Herne}, D. and {Hewitt}, J.~N. and {Johnston-Hollitt}, M. and {Kaplan}, D.~L. and {Kasper}, J.~C. and {Kincaid}, B.~B. and {Koenig}, R. and {Kratzenberg}, E. and {Lynch}, M.~J. and {Mckinley}, B. and {Mcwhirter}, S.~R. and {Morgan}, E. and {Oberoi}, D. and {Pathikulangara}, J. and {Prabu}, T. and {Remillard}, R.~A. and {Rogers}, A.~E.~E. and {Roshi}, A. and {Salah}, J.~E. and {Sault}, R.~J. and {Udaya-Shankar}, N. and {Schlagenhaufer}, F. and {Srivani}, K.~S. and {Stevens}, J. and {Subrahmanyan}, R. and {Waterson}, M. and {Webster}, R.~L. and {Whitney}, A.~R. and {Williams}, A. and {Williams}, C.~L. and {Wyithe}, J.~S.~B.},
        title = "{The Murchison Widefield Array: The Square Kilometre Array Precursor at Low Radio Frequencies}",
      journal = {\pasa},
         year = 2013,
        month = jan,
       volume = {30},
          eid = {e007},
        pages = {e007},
          doi = {10.1017/pasa.2012.007},
archivePrefix = {arXiv},
       eprint = {1206.6945},
 primaryClass = {astro-ph.IM},
       adsurl = {https://ui.adsabs.harvard.edu/abs/2013PASA...30....7T}
}

@article{LOFAR,
	author = {{van Haarlem, M. P.} and {Wise, M. W.} and {Gunst, A. W.} and {Heald, G.} and {McKean, J. P.} and {Hessels, J. W. T.} and {de Bruyn, A. G.} and {Nijboer, R.} and {Swinbank, J.} and {Fallows, R.} and {Brentjens, M.} and {Nelles, A.} and {Beck, R.} and {Falcke, H.} and {Fender, R.} and {Hörandel, J.} and {Koopmans, L. V. E.} and {Mann, G.} and {Miley, G.} and {Röttgering, H.} and {Stappers, B. W.} and {Wijers, R. A. M. J.} and {Zaroubi, S.} and {van den Akker, M.} and {Alexov, A.} and {Anderson, J.} and {Anderson, K.} and {van Ardenne, A.} and {Arts, M.} and {Asgekar, A.} and {Avruch, I. M.} and {Batejat, F.} and {Bähren, L.} and {Bell, M. E.} and {Bell, M. R.} and {van Bemmel, I.} and {Bennema, P.} and {Bentum, M. J.} and {Bernardi, G.} and {Best, P.} and {Bîrzan, L.} and {Bonafede, A.} and {Boonstra, A.-J.} and {Braun, R.} and {Bregman, J.} and {Breitling, F.} and {van de Brink, R. H.} and {Broderick, J.} and {Broekema, P. C.} and {Brouw, W. N.} and {Brüggen, M.} and {Butcher, H. R.} and {van Cappellen, W.} and {Ciardi, B.} and {Coenen, T.} and {Conway, J.} and {Coolen, A.} and {Corstanje, A.} and {Damstra, S.} and {Davies, O.} and {Deller, A. T.} and {Dettmar, R.-J.} and {van Diepen, G.} and {Dijkstra, K.} and {Donker, P.} and {Doorduin, A.} and {Dromer, J.} and {Drost, M.} and {van Duin, A.} and {Eislöffel, J.} and {van Enst, J.} and {Ferrari, C.} and {Frieswijk, W.} and {Gankema, H.} and {Garrett, M. A.} and {de Gasperin, F.} and {Gerbers, M.} and {de Geus, E.} and {Grießmeier, J.-M.} and {Grit, T.} and {Gruppen, P.} and {Hamaker, J. P.} and {Hassall, T.} and {Hoeft, M.} and {Holties, H. A.} and {Horneffer, A.} and {van der Horst, A.} and {van Houwelingen, A.} and {Huijgen, A.} and {Iacobelli, M.} and {Intema, H.} and {Jackson, N.} and {Jelic, V.} and {de Jong, A.} and {Juette, E.} and {Kant, D.} and {Karastergiou, A.} and {Koers, A.} and {Kollen, H.} and {Kondratiev, V. I.} and {Kooistra, E.} and {Koopman, Y.} and {Koster, A.} and {Kuniyoshi, M.} and {Kramer, M.} and {Kuper, G.} and {Lambropoulos, P.} and {Law, C.} and {van Leeuwen, J.} and {Lemaitre, J.} and {Loose, M.} and {Maat, P.} and {Macario, G.} and {Markoff, S.} and {Masters, J.} and {McFadden, R. A.} and {McKay-Bukowski, D.} and {Meijering, H.} and {Meulman, H.} and {Mevius, M.} and {Middelberg, E.} and {Millenaar, R.} and {Miller-Jones, J. C. A.} and {Mohan, R. N.} and {Mol, J. D.} and {Morawietz, J.} and {Morganti, R.} and {Mulcahy, D. D.} and {Mulder, E.} and {Munk, H.} and {Nieuwenhuis, L.} and {van Nieuwpoort, R.} and {Noordam, J. E.} and {Norden, M.} and {Noutsos, A.} and {Offringa, A. R.} and {Olofsson, H.} and {Omar, A.} and {Orrú, E.} and {Overeem, R.} and {Paas, H.} and {Pandey-Pommier, M.} and {Pandey, V. N.} and {Pizzo, R.} and {Polatidis, A.} and {Rafferty, D.} and {Rawlings, S.} and {Reich, W.} and {de Reijer, J.-P.} and {Reitsma, J.} and {Renting, G. A.} and {Riemers, P.} and {Rol, E.} and {Romein, J. W.} and {Roosjen, J.} and {Ruiter, M.} and {Scaife, A.} and {van der Schaaf, K.} and {Scheers, B.} and {Schellart, P.} and {Schoenmakers, A.} and {Schoonderbeek, G.} and {Serylak, M.} and {Shulevski, A.} and {Sluman, J.} and {Smirnov, O.} and {Sobey, C.} and {Spreeuw, H.} and {Steinmetz, M.} and {Sterks, C. G. M.} and {Stiepel, H.-J.} and {Stuurwold, K.} and {Tagger, M.} and {Tang, Y.} and {Tasse, C.} and {Thomas, I.} and {Thoudam, S.} and {Toribio, M. C.} and {van der Tol, B.} and {Usov, O.} and {van Veelen, M.} and {van der Veen, A.-J.} and {ter Veen, S.} and {Verbiest, J. P. W.} and {Vermeulen, R.} and {Vermaas, N.} and {Vocks, C.} and {Vogt, C.} and {de Vos, M.} and {van der Wal, E.} and {van Weeren, R.} and {Weggemans, H.} and {Weltevrede, P.} and {White, S.} and {Wijnholds, S. J.} and {Wilhelmsson, T.} and {Wucknitz, O.} and {Yatawatta, S.} and {Zarka, P.} and {Zensus, A.} and {van Zwieten, J.}},
	title = {LOFAR: The LOw-Frequency ARray},
	DOI= "10.1051/0004-6361/201220873",
	url= "https://doi.org/10.1051/0004-6361/201220873",
	journal = {A\&A},
	year = 2013,
	volume = 556,
	pages = "A2",
	month = "",
}

@article{DeBoer_2017,
doi = {10.1088/1538-3873/129/974/045001},
url = {https://dx.doi.org/10.1088/1538-3873/129/974/045001},
year = {2017},
month = {mar},
publisher = {The Astronomical Society of the Pacific},
volume = {129},
number = {974},
pages = {045001},
author = {David R. DeBoer and Aaron R. Parsons and James E. Aguirre and Paul Alexander and Zaki S. Ali and Adam P. Beardsley and Gianni Bernardi and Judd D. Bowman and Richard F. Bradley and Chris L. Carilli and Carina Cheng and Eloy de Lera Acedo and Joshua S. Dillon and Aaron Ewall-Wice and Gcobisa Fadana and Nicolas Fagnoni and Randall Fritz and Steve R. Furlanetto and Brian Glendenning and Bradley Greig and Jasper Grobbelaar and Bryna J. Hazelton and Jacqueline N. Hewitt and Jack Hickish and Daniel C. Jacobs and Austin Julius and MacCalvin Kariseb and Saul A. Kohn and Telalo Lekalake and Adrian Liu and Anita Loots and David MacMahon and Lourence Malan and Cresshim Malgas and Matthys Maree and Zachary Martinot and Nathan Mathison and Eunice Matsetela and Andrei Mesinger and Miguel F. Morales and Abraham R. Neben and Nipanjana Patra and Samantha Pieterse and Jonathan C. Pober and Nima Razavi-Ghods and Jon Ringuette and James Robnett and Kathryn Rosie and Raddwine Sell and Craig Smith and Angelo Syce and Max Tegmark and Nithyanandan Thyagarajan and Peter K. G. Williams and Haoxuan Zheng},
title = {Hydrogen Epoch of Reionization Array (HERA)},
journal = {PASP}
}

@article{Abdurashidova_2022,
   title={First Results from HERA Phase I: Upper Limits on the Epoch of Reionization 21 cm Power Spectrum},
   volume={925},
   ISSN={1538-4357},
   url={http://dx.doi.org/10.3847/1538-4357/ac1c78},
   DOI={10.3847/1538-4357/ac1c78},
   number={2},
   journal={ApJ},
   publisher={American Astronomical Society},
   author={Abdurashidova, Zara and Aguirre, James E. and Alexander, Paul and Ali, Zaki S. and Balfour, Yanga and Beardsley, Adam P. and Bernardi, Gianni and Billings, Tashalee S. and Bowman, Judd D. and Bradley, Richard F. and Bull, Philip and Burba, Jacob and Carey, Steve and Carilli, Chris L. and Cheng, Carina and DeBoer, David R. and Dexter, Matt and de Lera Acedo, Eloy and Dibblee-Barkman, Taylor and Dillon, Joshua S. and Ely, John and Ewall-Wice, Aaron and Fagnoni, Nicolas and Fritz, Randall and Furlanetto, Steven R. and Gale-Sides, Kingsley and Glendenning, Brian and Gorthi, Deepthi and Greig, Bradley and Grobbelaar, Jasper and Halday, Ziyaad and Hazelton, Bryna J. and Hewitt, Jacqueline N. and Hickish, Jack and Jacobs, Daniel C. and Julius, Austin and Kern, Nicholas S. and Kerrigan, Joshua and Kittiwisit, Piyanat and Kohn, Saul A. and Kolopanis, Matthew and Lanman, Adam and La Plante, Paul and Lekalake, Telalo and Lewis, David and Liu, Adrian and MacMahon, David and Malan, Lourence and Malgas, Cresshim and Maree, Matthys and Martinot, Zachary E. and Matsetela, Eunice and Mesinger, Andrei and Molewa, Mathakane and Morales, Miguel F. and Mosiane, Tshegofalang and Murray, Steven G. and Neben, Abraham R. and Nikolic, Bojan and Nunhokee, Chuneeta D. and Parsons, Aaron R. and Patra, Nipanjana and Pascua, Robert and Pieterse, Samantha and Pober, Jonathan C. and Razavi-Ghods, Nima and Ringuette, Jon and Robnett, James and Rosie, Kathryn and Sims, Peter and Singh, Saurabh and Smith, Craig and Syce, Angelo and Thyagarajan, Nithyanandan and Williams, Peter K. G. and Zheng, Haoxuan},
   year={2022},
   month=feb, pages={221} }

@article{Abdurashidova_2023,
   title={Improved Constraints on the 21 cm EoR Power Spectrum and the X-Ray Heating of the IGM with HERA Phase I Observations},
   volume={945},
   ISSN={1538-4357},
   url={http://dx.doi.org/10.3847/1538-4357/acaf50},
   DOI={10.3847/1538-4357/acaf50},
   number={2},
   journal={ApJ},
   publisher={American Astronomical Society},
   author={Abdurashidova, The HERA Collaboration: Zara and Adams, Tyrone and Aguirre, James E. and Alexander, Paul and Ali, Zaki S. and Baartman, Rushelle and Balfour, Yanga and Barkana, Rennan and Beardsley, Adam P. and Bernardi, Gianni and Billings, Tashalee S. and Bowman, Judd D. and Bradley, Richard F. and Breitman, Daniela and Bull, Philip and Burba, Jacob and Carey, Steve and Carilli, Chris L. and Cheng, Carina and Choudhuri, Samir and DeBoer, David R. and de Lera Acedo, Eloy and Dexter, Matt and Dillon, Joshua S. and Ely, John and Ewall-Wice, Aaron and Fagnoni, Nicolas and Fialkov, Anastasia and Fritz, Randall and Furlanetto, Steven R. and Gale-Sides, Kingsley and Garsden, Hugh and Glendenning, Brian and Gorce, Adélie and Gorthi, Deepthi and Greig, Bradley and Grobbelaar, Jasper and Halday, Ziyaad and Hazelton, Bryna J. and Heimersheim, Stefan and Hewitt, Jacqueline N. and Hickish, Jack and Jacobs, Daniel C. and Julius, Austin and Kern, Nicholas S. and Kerrigan, Joshua and Kittiwisit, Piyanat and Kohn, Saul A. and Kolopanis, Matthew and Lanman, Adam and La Plante, Paul and Lewis, David and Liu, Adrian and Loots, Anita and Ma, Yin-Zhe and MacMahon, David H. E. and Malan, Lourence and Malgas, Keith and Malgas, Cresshim and Maree, Matthys and Marero, Bradley and Martinot, Zachary E. and McBride, Lisa and Mesinger, Andrei and Mirocha, Jordan and Molewa, Mathakane and Morales, Miguel F. and Mosiane, Tshegofalang and Muñoz, Julian B. and Murray, Steven G. and Nagpal, Vighnesh and Neben, Abraham R. and Nikolic, Bojan and Nunhokee, Chuneeta D. and Nuwegeld, Hans and Parsons, Aaron R. and Pascua, Robert and Patra, Nipanjana and Pieterse, Samantha and Qin, Yuxiang and Razavi-Ghods, Nima and Robnett, James and Rosie, Kathryn and Santos, Mario G. and Sims, Peter and Singh, Saurabh and Smith, Craig and Swarts, Hilton and Tan, Jianrong and Thyagarajan, Nithyanandan and Wilensky, Michael J. and Williams, Peter K. G. and van Wyngaarden, Pieter and Zheng, Haoxuan},
   year={2023},
   month=mar, pages={124} }

@ARTICLE{Ceccotti_2025,
       author = {{Ceccotti}, E. and {Offringa}, A.~R. and {Mertens}, F.~G. and {Koopmans}, L.~V.~E. and {Munshi}, S. and {Chege}, J.~K. and {Acharya}, A. and {Brackenhoff}, S.~A. and {Chapman}, E. and {Ciardi}, B. and {Ghara}, R. and {Ghosh}, S. and {Giri}, S.~K. and {H{\"o}fer}, C. and {Hothi}, I. and {Mellema}, G. and {Mevius}, M. and {Pandey}, V.~N. and {Zaroubi}, S.},
        title = "{First upper limits on the 21-cm signal power spectrum of neutral hydrogen at z = 9.16 from the LOFAR 3C 196 field}",
      journal = {\mnras},
         year = 2025,
        month = nov,
       volume = {544},
       number = {1},
        pages = {1255-1283},
          doi = {10.1093/mnras/staf1629},
archivePrefix = {arXiv},
       eprint = {2504.18534},
 primaryClass = {astro-ph.CO},
       adsurl = {https://ui.adsabs.harvard.edu/abs/2025MNRAS.544.1255C}
}

@ARTICLE{Trott_2020,
       author = {{Trott}, Cathryn M. and {Jordan}, C.~H. and {Midgley}, S. and {Barry}, N. and {Greig}, B. and {Pindor}, B. and {Cook}, J.~H. and {Sleap}, G. and {Tingay}, S.~J. and {Ung}, D. and {Hancock}, P. and {Williams}, A. and {Bowman}, J. and {Byrne}, R. and {Chokshi}, A. and {Hazelton}, B.~J. and {Hasegawa}, K. and {Jacobs}, D. and {Joseph}, R.~C. and {Li}, W. and {Line}, J.~L.~B. and {Lynch}, C. and {McKinley}, B. and {Mitchell}, D.~A. and {Morales}, M.~F. and {Ouchi}, M. and {Pober}, J.~C. and {Rahimi}, M. and {Takahashi}, K. and {Wayth}, R.~B. and {Webster}, R.~L. and {Wilensky}, M. and {Wyithe}, J.~S.~B. and {Yoshiura}, S. and {Zhang}, Z. and {Zheng}, Q.},
        title = "{Deep multiredshift limits on Epoch of Reionization 21 cm power spectra from four seasons of Murchison Widefield Array observations}",
      journal = {\mnras},
         year = 2020,
        month = apr,
       volume = {493},
       number = {4},
        pages = {4711-4727},
          doi = {10.1093/mnras/staa414},
archivePrefix = {arXiv},
       eprint = {2002.02575},
 primaryClass = {astro-ph.CO},
       adsurl = {https://ui.adsabs.harvard.edu/abs/2020MNRAS.493.4711T}
}

@ARTICLE{Mertens_2020,
       author = {{Mertens}, F.~G. and {Mevius}, M. and {Koopmans}, L.~V.~E. and {Offringa}, A.~R. and {Mellema}, G. and {Zaroubi}, S. and {Brentjens}, M.~A. and {Gan}, H. and {Gehlot}, B.~K. and {Pandey}, V.~N. and {Sardarabadi}, A.~M. and {Vedantham}, H.~K. and {Yatawatta}, S. and {Asad}, K.~M.~B. and {Ciardi}, B. and {Chapman}, E. and {Gazagnes}, S. and {Ghara}, R. and {Ghosh}, A. and {Giri}, S.~K. and {Iliev}, I.~T. and {Jeli{\'c}}, V. and {Kooistra}, R. and {Mondal}, R. and {Schaye}, J. and {Silva}, M.~B.},
        title = "{Improved upper limits on the 21 cm signal power spectrum of neutral hydrogen at z {\ensuremath{\approx}} 9.1 from LOFAR}",
      journal = {\mnras},
         year = 2020,
        month = apr,
       volume = {493},
       number = {2},
        pages = {1662-1685},
          doi = {10.1093/mnras/staa327},
archivePrefix = {arXiv},
       eprint = {2002.07196},
 primaryClass = {astro-ph.CO},
       adsurl = {https://ui.adsabs.harvard.edu/abs/2020MNRAS.493.1662M}
}

@ARTICLE{Yoshiura_2021,
       author = {{Yoshiura}, S. and {Pindor}, B. and {Line}, J.~L.~B. and {Barry}, N. and {Trott}, C.~M. and {Beardsley}, A. and {Bowman}, J. and {Byrne}, R. and {Chokshi}, A. and {Hazelton}, B.~J. and {Hasegawa}, K. and {Howard}, E. and {Greig}, B. and {Jacobs}, D. and {Jordan}, C.~H. and {Joseph}, R. and {Kolopanis}, M. and {Lynch}, C. and {McKinley}, B. and {Mitchell}, D.~A. and {Morales}, M.~F. and {Murray}, S.~G. and {Pober}, J.~C. and {Rahimi}, M. and {Takahashi}, K. and {Tingay}, S.~J. and {Wayth}, R.~B. and {Webster}, R.~L. and {Wilensky}, M. and {Wyithe}, J.~S.~B. and {Zhang}, Z. and {Zheng}, Q.},
        title = "{A new MWA limit on the 21 cm power spectrum at redshifts  13-17}",
      journal = {\mnras},
         year = 2021,
        month = aug,
       volume = {505},
       number = {4},
        pages = {4775-4790},
          doi = {10.1093/mnras/stab1560},
archivePrefix = {arXiv},
       eprint = {2105.12888},
 primaryClass = {astro-ph.CO},
       adsurl = {https://ui.adsabs.harvard.edu/abs/2021MNRAS.505.4775Y}
}

@ARTICLE{Liu_2020,
       author = {{Liu}, Xue-Wen and {Heneka}, Caroline and {Amendola}, Luca},
        title = "{Constraining coupled quintessence with the 21 cm signal}",
      journal = {\jcap},
         year = 2020,
        month = may,
       volume = {2020},
       number = {5},
          eid = {038},
        pages = {038},
          doi = {10.1088/1475-7516/2020/05/038},
archivePrefix = {arXiv},
       eprint = {1910.02763},
 primaryClass = {astro-ph.CO},
       adsurl = {https://ui.adsabs.harvard.edu/abs/2020JCAP...05..038L}
}

@ARTICLE{Pietschke_2025,
       author = {{Pietschke}, Yannic and {Heneka}, Caroline and {Schlenker}, Tom and {Ore}, Ayodele and {Schosser}, Benedikt},
        title = "{Direct reconstruction of the Reionization history from 21cm 2D Power Spectra}",
      journal = {JCAP},
         year = 2025,
        month = oct,
       volume = {2025},
       number = {10},
          eid = {039},
        pages = {039},
          doi = {10.1088/1475-7516/2025/10/039},
archivePrefix = {arXiv},
       eprint = {2506.19925},
 primaryClass = {astro-ph.CO},
       adsurl = {https://ui.adsabs.harvard.edu/abs/2025JCAP...10..039P}
}

@Article{Ore_2025,
	title={{SKATR: A self-supervised summary transformer for SKA}},
	author={Ayodele Ore and Caroline Heneka and Tilman Plehn},
	journal={SciPost Phys.},
	volume={18},
	pages={155},
	year={2025},
	publisher={SciPost},
	doi={10.21468/SciPostPhys.18.5.155},
	url={https://scipost.org/10.21468/SciPostPhys.18.5.155},
}

@ARTICLE{Lemos_2023,
       author = {{Lemos}, Pablo and {Coogan}, Adam and {Hezaveh}, Yashar and {Perreault-Levasseur}, Laurence},
        title = "{Sampling-Based Accuracy Testing of Posterior Estimators for General Inference}",
      journal = {40th International Conference on Machine Learning},
         year = 2023,
        month = jan,
       volume = {202},
        pages = {19256-19273},
          doi = {10.48550/arXiv.2302.03026},
archivePrefix = {arXiv},
       eprint = {2302.03026},
 primaryClass = {stat.ML},
       adsurl = {https://ui.adsabs.harvard.edu/abs/2023PMLR..20219256L}
}

@article{21cmfast, doi = {10.21105/joss.02582}, url = {https://doi.org/10.21105/joss.02582}, year = {2020}, publisher = {The Open Journal}, volume = {5}, number = {54}, pages = {2582}, author = {Steven G. Murray and Bradley Greig and Andrei Mesinger and Julian B. Muñoz and Yuxiang Qin and Jaehong Park and Catherine A. Watkinson}, title = {21cmFAST v3: A Python-integrated C code for generating 3D realizations of the cosmic 21cm signal.}, journal = {JOSS} }

@article{21cmfast11,
	title = {21cmfast: a fast, seminumerical simulation of the high-redshift 21-cm signal},
	volume = {411},
	issn = {0035-8711},
	url = {https://doi.org/10.1111/j.1365-2966.2010.17731.x},
	doi = {10.1111/j.1365-2966.2010.17731.x},
	number = {2},
	journal = {\mnras},
	author = {Mesinger, Andrei and Furlanetto, Steven and Cen, Renyue},
	month = feb,
	year = {2011},
	pages = {955--972},
}

@ARTICLE{Alsing_2019,
       author = {{Alsing}, Justin and {Charnock}, Tom and {Feeney}, Stephen and {Wandelt}, Benjamin},
        title = "{Fast likelihood-free cosmology with neural density estimators and active learning}",
      journal = {\mnras},
         year = 2019,
        month = sep,
       volume = {488},
       number = {3},
        pages = {4440-4458},
          doi = {10.1093/mnras/stz1960},
archivePrefix = {arXiv},
       eprint = {1903.00007},
 primaryClass = {astro-ph.CO},
       adsurl = {https://ui.adsabs.harvard.edu/abs/2019MNRAS.488.4440A}
}

@ARTICLE{Cole_2022,
       author = {{Cole}, Alex and {Miller}, Benjamin K. and {Witte}, Samuel J. and {Cai}, Maxwell X. and {Grootes}, Meiert W. and {Nattino}, Francesco and {Weniger}, Christoph},
        title = "{Fast and credible likelihood-free cosmology with truncated marginal neural ratio estimation}",
      journal = {JCAP},
         year = 2022,
        month = sep,
       volume = {2022},
       number = {9},
          eid = {004},
        pages = {004},
          doi = {10.1088/1475-7516/2022/09/004},
archivePrefix = {arXiv},
       eprint = {2111.08030},
 primaryClass = {astro-ph.CO},
       adsurl = {https://ui.adsabs.harvard.edu/abs/2022JCAP...09..004C}
}

@ARTICLE{Saxena_2024,
       author = {{Saxena}, Anchal and {Meerburg}, P. Daniel and {Weniger}, Christoph and {Acedo}, Eloy de Lera and {Handley}, Will},
        title = "{Simulation-based inference of the sky-averaged 21-cm signal from CD-EoR with REACH}",
      journal = {RASTI},
         year = 2024,
        month = jan,
       volume = {3},
       number = {1},
        pages = {724-736},
          doi = {10.1093/rasti/rzae047},
archivePrefix = {arXiv},
       eprint = {2403.14618},
 primaryClass = {astro-ph.CO},
       adsurl = {https://ui.adsabs.harvard.edu/abs/2024RASTI...3..724S}
}

@ARTICLE{Prelogovic_2024,
       author = {{Prelogovi{\'c}}, David and {Mesinger}, Andrei},
        title = "{How informative are summaries of the cosmic 21 cm signal?}",
      journal = {\aap},
         year = 2024,
        month = aug,
       volume = {688},
          eid = {A199},
        pages = {A199},
          doi = {10.1051/0004-6361/202449309},
archivePrefix = {arXiv},
       eprint = {2401.12277},
 primaryClass = {astro-ph.CO},
       adsurl = {https://ui.adsabs.harvard.edu/abs/2024A&A...688A.199P}
}

@ARTICLE{Schosser_2025,
       author = {{Schosser}, Benedikt and {Heneka}, Caroline and {Plehn}, Tilman},
        title = "{Optimal, fast, and robust inference of reionization-era cosmology with the 21cmPIE-INN}",
      journal = {SciPost Physics Core},
         year = 2025,
        month = apr,
       volume = {8},
       number = {2},
          eid = {037},
        pages = {037},
          doi = {10.21468/SciPostPhysCore.8.2.037},
archivePrefix = {arXiv},
       eprint = {2401.04174},
 primaryClass = {astro-ph.CO},
       adsurl = {https://ui.adsabs.harvard.edu/abs/2025ScPC....8...37S}
}

@ARTICLE{Schosser_2026,
       author = {{Schosser}, Benedikt and {Heneka}, Caroline and {Sch{\"a}fer}, Bj{\"o}rn Malte},
        title = "{Starobinsky in stereo: SKA-CMB synergy in SBI}",
      journal = {JCAP},
         year = 2026,
        month = feb,
       volume = {2026},
       number = {2},
          eid = {086},
        pages = {086},
          doi = {10.1088/1475-7516/2026/02/086},
archivePrefix = {arXiv},
       eprint = {2508.10094},
 primaryClass = {astro-ph.CO},
       adsurl = {https://ui.adsabs.harvard.edu/abs/2026JCAP...02..086S}
}

@article{Park_2019,
   title={Inferring the astrophysics of reionization and cosmic dawn from galaxy luminosity functions and the 21-cm signal},
   volume={484},
   ISSN={1365-2966},
   url={http://dx.doi.org/10.1093/mnras/stz032},
   DOI={10.1093/mnras/stz032},
   number={1},
   journal={\mnras},
   publisher={Oxford University Press (OUP)},
   author={Park, Jaehong and Mesinger, Andrei and Greig, Bradley and Gillet, Nicolas},
   year={2019},
   month=jan, pages={933–949} }

@ARTICLE{Meriot_24,
       author = {{Meriot}, R. and {Semelin}, B.},
        title = "{The LORELI database: 21 cm signal inference with 3D radiative hydrodynamics simulations}",
      journal = {\aap},
         year = 2024,
        month = mar,
       volume = {683},
          eid = {A24},
        pages = {A24},
          doi = {10.1051/0004-6361/202347591},
archivePrefix = {arXiv},
       eprint = {2310.02684},
 primaryClass = {astro-ph.CO},
       adsurl = {https://ui.adsabs.harvard.edu/abs/2024A&A...683A..24M}
}

@ARTICLE{Meriot_2025,
       author = {{Meriot}, R. and {Semelin}, B. and {Cornu}, D.},
        title = "{Comparison of Bayesian inference methods using the LORELI II database of hydro-radiative simulations of the 21-cm signal}",
      journal = {\aap},
         year = 2025,
        month = jun,
       volume = {698},
          eid = {A80},
        pages = {A80},
          doi = {10.1051/0004-6361/202452901},
archivePrefix = {arXiv},
       eprint = {2411.03093},
 primaryClass = {astro-ph.CO},
       adsurl = {https://ui.adsabs.harvard.edu/abs/2025A&A...698A..80M}
}

@ARTICLE{Semelin_2023,
       author = {{Semelin}, B. and {M{\'e}riot}, R. and {Mertens}, F. and {Koopmans}, L.~V.~E. and {Aubert}, D. and {Barkana}, R. and {Fialkov}, A. and {Munshi}, S. and {Ocvirk}, P.},
        title = "{Accurate modelling of the Lyman-{\ensuremath{\alpha}} coupling for the 21-cm signal, observability with NenuFAR, and SKA}",
      journal = {\aap},
         year = 2023,
        month = apr,
       volume = {672},
          eid = {A162},
        pages = {A162},
          doi = {10.1051/0004-6361/202244722},
archivePrefix = {arXiv},
       eprint = {2301.13178},
 primaryClass = {astro-ph.CO},
       adsurl = {https://ui.adsabs.harvard.edu/abs/2023A&A...672A.162S}
}

@article{Oesch_2018,
doi = {10.3847/1538-4357/aab03f},
url = {https://doi.org/10.3847/1538-4357/aab03f},
year = {2018},
month = {mar},
publisher = {The American Astronomical Society},
volume = {855},
number = {2},
pages = {105},
author = {Oesch, P. A. and Bouwens, R. J. and Illingworth, G. D. and Labbé, I. and Stefanon, M.},
title = {The Dearth of z ∼ 10 Galaxies in All HST Legacy Fields—The Rapid Evolution of the Galaxy Population in the First 500 Myr*},
journal = {The Astrophysical Journal}
}

@ARTICLE{McLeod_2016,
       author = {{McLeod}, D.~J. and {McLure}, R.~J. and {Dunlop}, J.~S.},
        title = "{The z = 9-10 galaxy population in the Hubble Frontier Fields and CLASH surveys: the z = 9 luminosity function and further evidence for a smooth decline in ultraviolet luminosity density at z{\ensuremath{\geq}} 8}",
      journal = {\mnras},
         year = 2016,
        month = jul,
       volume = {459},
       number = {4},
        pages = {3812-3824},
          doi = {10.1093/mnras/stw904},
archivePrefix = {arXiv},
       eprint = {1602.05199},
 primaryClass = {astro-ph.GA},
       adsurl = {https://ui.adsabs.harvard.edu/abs/2016MNRAS.459.3812M}
}

@ARTICLE{Bouwens_2016,
       author = {{Bouwens}, Rychard J. and {Aravena}, Manuel and {Decarli}, Roberto and {Walter}, Fabian and {da Cunha}, Elisabete and {Labb{\'e}}, Ivo and {Bauer}, Franz E. and {Bertoldi}, Frank and {Carilli}, Chris and {Chapman}, Scott and {Daddi}, Emanuele and {Hodge}, Jacqueline and {Ivison}, Rob J. and {Karim}, Alex and {Le Fevre}, Olivier and {Magnelli}, Benjamin and {Ota}, Kazuaki and {Riechers}, Dominik and {Smail}, Ian R. and {van der Werf}, Paul and {Weiss}, Axel and {Cox}, Pierre and {Elbaz}, David and {Gonzalez-Lopez}, Jorge and {Infante}, Leopoldo and {Oesch}, Pascal and {Wagg}, Jeff and {Wilkins}, Steve},
        title = "{ALMA Spectroscopic Survey in the Hubble Ultra Deep Field: The Infrared Excess of UV-Selected z = 2-10 Galaxies as a Function of UV-Continuum Slope and Stellar Mass}",
      journal = {\apj},
         year = 2016,
        month = dec,
       volume = {833},
       number = {1},
          eid = {72},
        pages = {72},
          doi = {10.3847/1538-4357/833/1/72},
archivePrefix = {arXiv},
       eprint = {1606.05280},
 primaryClass = {astro-ph.GA},
       adsurl = {https://ui.adsabs.harvard.edu/abs/2016ApJ...833...72B}
}

@article{Planck2018,
    author = "Aghanim, N. and others",
    collaboration = "Planck",
    title = "{Planck 2018 results. VI. Cosmological parameters}",
    eprint = "1807.06209",
    archivePrefix = "arXiv",
    primaryClass = "astro-ph.CO",
    doi = "10.1051/0004-6361/201833910",
    journal = "\aap",
    volume = "641",
    pages = "A6",
    year = "2020",
    note = "[Erratum: A\&A 652, C4 (2021)]"
}

@ARTICLE{Cranmer_2020,
       author = {{Cranmer}, Kyle and {Brehmer}, Johann and {Louppe}, Gilles},
        title = "{The frontier of simulation-based inference}",
      journal = {Proceedings of the National Academy of Science},
         year = 2020,
        month = dec,
       volume = {117},
       number = {48},
        pages = {30055-30062},
          doi = {10.1073/pnas.1912789117},
archivePrefix = {arXiv},
       eprint = {1911.01429},
 primaryClass = {stat.ML},
       adsurl = {https://ui.adsabs.harvard.edu/abs/2020PNAS..11730055C}
}

@ARTICLE{Sooknunan_2024,
       author = {{Sooknunan}, Kimeel and {Chapman}, Emma and {Conaboy}, Luke and {Mortlock}, Daniel and {Pritchard}, Jonathan},
        title = "{Reproducibility of machine learning analyses of 21 cm reionization maps}",
      journal = {\mnras},
         note = {Submitted},
         year = 2024,
        month = dec,
          eid = {arXiv:2412.15893},
        pages = {arXiv:2412.15893},
          doi = {10.48550/arXiv.2412.15893},
archivePrefix = {arXiv},
       eprint = {2412.15893},
 primaryClass = {astro-ph.CO},
       adsurl = {https://ui.adsabs.harvard.edu/abs/2024arXiv241215893S}
}

@ARTICLE{Cannon_2022,
       author = {{Cannon}, Patrick and {Ward}, Daniel and {Schmon}, Sebastian M.},
        title = "{Investigating the Impact of Model Misspecification in Neural Simulation-based Inference}",
      journal = {arXiv e-prints},
         year = 2022,
        month = sep,
          eid = {arXiv:2209.01845},
        pages = {arXiv:2209.01845},
          doi = {10.48550/arXiv.2209.01845},
archivePrefix = {arXiv},
       eprint = {2209.01845},
 primaryClass = {stat.ML},
       adsurl = {https://ui.adsabs.harvard.edu/abs/2022arXiv220901845C}
}

@InProceedings{Assran_2023,
    author    = {Assran, Mahmoud and Duval, Quentin and Misra, Ishan and Bojanowski, Piotr and Vincent, Pascal and Rabbat, Michael and LeCun, Yann and Ballas, Nicolas},
    title     = {Self-Supervised Learning From Images With a Joint-Embedding Predictive Architecture},
    booktitle = {Proceedings of the IEEE/CVF Conference on Computer Vision and Pattern Recognition (CVPR)},
    month     = {June},
    year      = {2023},
    pages     = {15619-15629}
}

@article{
Bardes_2024,
title={Revisiting Feature Prediction for Learning Visual Representations from Video},
author={Adrien Bardes and Quentin Garrido and Jean Ponce and Xinlei Chen and Michael Rabbat and Yann LeCun and Mido Assran and Nicolas Ballas},
journal={Transactions on Machine Learning Research},
issn={2835-8856},
year={2024},
url={https://openreview.net/forum?id=QaCCuDfBk2},
note={Featured Certification}
}

@inproceedings{
Dosovitskiy_2021,
title={An Image is Worth 16x16 Words: Transformers for Image Recognition at Scale},
author={Alexey Dosovitskiy and Lucas Beyer and Alexander Kolesnikov and Dirk Weissenborn and Xiaohua Zhai and Thomas Unterthiner and Mostafa Dehghani and Matthias Minderer and Georg Heigold and Sylvain Gelly and Jakob Uszkoreit and Neil Houlsby},
booktitle={International Conference on Learning Representations},
year={2021},
url={https://openreview.net/forum?id=YicbFdNTTy}
}

@inproceedings{
Sui_2024,
title={Self-supervised Representation Learning from Random Data Projectors},
author={Yi Sui and Tongzi Wu and Jesse C. Cresswell and Ga Wu and George Stein and Xiao Shi Huang and Xiaochen Zhang and Maksims Volkovs},
booktitle={The Twelfth International Conference on Learning Representations},
year={2024},
url={https://openreview.net/forum?id=EpYnZpDpsQ}
}

@article{21cmsense13,
doi = {10.1088/0004-6256/145/3/65},
url = {https://dx.doi.org/10.1088/0004-6256/145/3/65},
year = {2013},
month = {jan},
publisher = {The American Astronomical Society},
volume = {145},
number = {3},
pages = {65},
author = {Jonathan C. Pober and Aaron R. Parsons and David R. DeBoer and Patrick McDonald and Matthew McQuinn and James E. Aguirre and Zaki Ali and Richard F. Bradley and Tzu-Ching Chang and Miguel F. Morales},
title = {THE BARYON ACOUSTIC OSCILLATION BROADBAND AND BROAD-BEAM ARRAY: DESIGN OVERVIEW AND SENSITIVITY FORECASTS},
journal = {\aj}
}

@article{21cmsense14,
doi = {10.1088/0004-637X/782/2/66},
url = {https://dx.doi.org/10.1088/0004-637X/782/2/66},
year = {2014},
month = {jan},
publisher = {The American Astronomical Society},
volume = {782},
number = {2},
pages = {66},
author = {Jonathan C. Pober and Adrian Liu and Joshua S. Dillon and James E. Aguirre and Judd D. Bowman and Richard F. Bradley and Chris L. Carilli and David R. DeBoer and Jacqueline N. Hewitt and Daniel C. Jacobs and Matthew McQuinn and Miguel F. Morales and Aaron R. Parsons and Max Tegmark and Dan J. Werthimer},
title = {WHAT NEXT-GENERATION 21 cm POWER SPECTRUM MEASUREMENTS CAN TEACH US ABOUT THE EPOCH OF REIONIZATION},
journal = {ApJ}
}

@inproceedings{Roncoli_2023,
    author = "Roncoli, Andrea and {\'C}iprijanovi{\'c}, Aleksandra and Voetberg, Maggie and Villaescusa-Navarro, Francisco and Nord, Brian",
    title = "{Domain Adaptive Graph Neural Networks for Constraining Cosmological Parameters Across Multiple Data Sets}",
    booktitle = "{37th Conference on Neural Information Processing Systems}",
    eprint = "2311.01588",
    archivePrefix = "arXiv",
    primaryClass = "astro-ph.CO",
    reportNumber = "FERMILAB-CONF-23-644-CSAID",
    month = "11",
    year = "2023"
}

@ARTICLE{Pierre_2026,
       author = {{Pierre}, S{\'e}bastien and {R{\'e}galdo-Saint Blancard}, Bruno and {Hahn}, ChangHoon and {Eickenberg}, Michael},
        title = "{Mitigating model misspecification in simulation-based inference for galaxy clustering}",
      journal = {\prd},
         year = 2026,
        month = feb,
       volume = {113},
       number = {4},
          eid = {043536},
        pages = {043536},
          doi = {10.1103/gypc-sqnx},
archivePrefix = {arXiv},
       eprint = {2507.03086},
 primaryClass = {astro-ph.CO},
       adsurl = {https://ui.adsabs.harvard.edu/abs/2026PhRvD.113d3536P}
}

@ARTICLE{Ntampaka_2025,
       author = {{Ntampaka}, Michelle and {Ciprijanovic}, A. and {Delgado}, Ana Maria and {Soltis}, John and {Wu}, John F. and {Yunus}, Mikaeel and {ZuHone}, John},
        title = "{The Importance of Being Adaptable: An Exploration of the Power and Limitations of Domain Adaptation for Simulation-Based Inference with Galaxy Clusters}",
      journal = {arXiv e-prints},
         year = 2025,
        month = oct,
        pages = {arXiv:2510.09748},
          doi = {10.48550/arXiv.2510.09748},
archivePrefix = {arXiv},
       eprint = {2510.09748},
 primaryClass = {astro-ph.IM},
       adsurl = {https://ui.adsabs.harvard.edu/abs/2025arXiv251009748N}
}

@INPROCEEDINGS{Lipman_2023,
       author = {{Lipman}, Yaron and {Chen}, Ricky T.~Q. and {Ben-Hamu}, Heli and {Nickel}, Maximilian and {Le}, Matt},
        title = "{Flow Matching for Generative Modeling}",
      journal = {ICLR},
         year = 2023,
      booktitle = {Proceedings of ICLR},
          doi = {10.48550/arXiv.2210.02747},
archivePrefix = {arXiv},
       eprint = {2210.02747}
}

@INPROCEEDINGS{Tong_2023,
       author = {{Tong}, Alexander and {Fatras}, Kilian and {Malkin}, Nikolay and {Huguet}, Guillaume and {Zhang}, Yanlei and {Rector-Brooks}, Jarrid and {Wolf}, Guy and {Bengio}, Yoshua},
        title = "{Improving and generalizing flow-based generative models with minibatch optimal transport}",
      journal = {TMLR},
      booktitle = {Proceedings of TMLR},
         year = 2024,
          url = {https://openreview.net/forum?id=CD9Snc73AW},
        issn = {2835-8856}
}

@ARTICLE{Talts_2018,
       author = {{Talts}, Sean and {Betancourt}, Michael and {Simpson}, Daniel and {Vehtari}, Aki and {Gelman}, Andrew},
        title = "{Validating Bayesian Inference Algorithms with Simulation-Based Calibration}",
      journal = {arXiv e-prints},
         year = 2018,
        month = apr,
          eid = {arXiv:1804.06788},
        pages = {arXiv:1804.06788},
          doi = {10.48550/arXiv.1804.06788},
archivePrefix = {arXiv},
       eprint = {1804.06788},
 primaryClass = {stat.ME},
       adsurl = {https://ui.adsabs.harvard.edu/abs/2018arXiv180406788T}
}

@ARTICLE{Parker_2024,
       author = {{Parker}, Liam and {Lanusse}, Francois and {Golkar}, Siavash and {Sarra}, Leopoldo and {Cranmer}, Miles and {Bietti}, Alberto and {Eickenberg}, Michael and {Krawezik}, Geraud and {McCabe}, Michael and {Morel}, Rudy and {Ohana}, Ruben and {Pettee}, Mariel and {R{\'e}galdo-Saint Blancard}, Bruno and {Cho}, Kyunghyun and {Ho}, Shirley and {Polymathic AI Collaboration}},
        title = "{AstroCLIP: a cross-modal foundation model for galaxies}",
      journal = {\mnras},
         year = 2024,
        month = jul,
       volume = {531},
       number = {4},
        pages = {4990-5011},
          doi = {10.1093/mnras/stae1450},
archivePrefix = {arXiv},
       eprint = {2310.03024},
 primaryClass = {astro-ph.IM},
       adsurl = {https://ui.adsabs.harvard.edu/abs/2024MNRAS.531.4990P}
}

@inproceedings{Smith_2024,
title={Astro{PT}: Scaling Large Observation Models for Astronomy},
author={Michael J. Smith and Ryan J. Roberts and Eirini Angeloudi and Marc Huertas-Company},
booktitle={ICML 2024 AI for Science Workshop},
year={2024},
url={https://openreview.net/forum?id=aOLuuLxqav}
}

@inproceedings{Parker_2025,
title={{AION}-1: Omnimodal Foundation Model for Astronomical Sciences},
author={Liam Holden Parker and Francois Lanusse and Jeff Shen and Ollie Liu and Tom Hehir and Leopoldo Sarra and Lucas Thibaut Meyer and Micah Bowles and Sebastian Wagner-Carena and Helen Qu and Siavash Golkar and Alberto Bietti and Hatim Bourfoune and Pierre Cornette and Keiya Hirashima and Geraud Krawezik and Ruben Ohana and Nicholas Lourie and Michael McCabe and Rudy Morel and Payel Mukhopadhyay and Mariel Pettee and Kyunghyun Cho and Miles Cranmer and Shirley Ho},
booktitle={The Thirty-ninth Annual Conference on Neural Information Processing Systems},
year={2025},
url={https://openreview.net/forum?id=6gJ2ZykQ5W}
}

@article{Hirling_2023,
    author = "Hirling, Patrick and Bianco, Michele and Giri, Sambit K. and Iliev, Ilian T. and Mellema, Garrelt and Kneib, Jean-Paul",
    title = "{pyC2Ray: A flexible and GPU-accelerated radiative transfer framework for simulating the cosmic epoch of reionization}",
    eprint = "2311.01492",
    archivePrefix = "arXiv",
    primaryClass = "astro-ph.CO",
    reportNumber = "NORDITA 2023-033",
    doi = "10.1016/j.ascom.2024.100861",
    journal = "Astron. Comput.",
    volume = "48",
    pages = "100861",
    year = "2024"
}

@ARTICLE{Saxena_2023,
       author = {{Saxena}, Anchal and {Cole}, Alex and {Gazagnes}, Simon and {Meerburg}, P. Daniel and {Weniger}, Christoph and {Witte}, Samuel J.},
        title = "{Constraining the X-ray heating and reionization using 21-cm power spectra with Marginal Neural Ratio Estimation}",
      journal = {\mnras},
         year = 2023,
        month = nov,
       volume = {525},
       number = {4},
        pages = {6097-6111},
          doi = {10.1093/mnras/stad2659},
archivePrefix = {arXiv},
       eprint = {2303.07339},
 primaryClass = {astro-ph.CO},
       adsurl = {https://ui.adsabs.harvard.edu/abs/2023MNRAS.525.6097S}
}

@ARTICLE{Greig_2024,
       author = {{Greig}, Bradley and {Prelogovi{\'c}}, David and {Qin}, Yuxiang and {Ting}, Yuan-Sen and {Mesinger}, Andrei},
        title = "{Inferring astrophysical parameters using the 2D cylindrical power spectrum from reionization}",
      journal = {\mnras},
         year = 2024,
        month = sep,
       volume = {533},
       number = {2},
        pages = {2530-2545},
          doi = {10.1093/mnras/stae1984},
archivePrefix = {arXiv},
       eprint = {2403.14060},
 primaryClass = {astro-ph.CO},
       adsurl = {https://ui.adsabs.harvard.edu/abs/2024MNRAS.533.2530G}
}

@ARTICLE{Cooper_2026,
       author = {{Cooper}, Nadia and {Norregaard}, Carina and {Meriot}, Romain and {Pritchard}, Jonathan R.},
        title = "{Simulation-based inference pipeline of the ionization history from the 2D 21 cm power spectrum}",
      journal = {\mnras},
         year = 2026,
        month = may,
       volume = {548},
       number = {2},
          eid = {stag577},
        pages = {stag577},
          doi = {10.1093/mnras/stag577},
archivePrefix = {arXiv},
       eprint = {2508.16329},
 primaryClass = {astro-ph.CO},
       adsurl = {https://ui.adsabs.harvard.edu/abs/2026MNRAS.548ag577C}
}

@ARTICLE{Pietschke_2026,
       author = {{Pietschke}, Yannic and {Hutter}, Anne and {Heneka}, Caroline},
        title = "{Constraining reionization morphology and source properties with 21 cm galaxy cross-correlation surveys}",
      journal = {\aap},
         year = 2026,
        month = jun,
       volume = {710},
          eid = {A113},
        pages = {A113},
          doi = {10.1051/0004-6361/202659152},
archivePrefix = {arXiv},
       eprint = {2601.18627},
 primaryClass = {astro-ph.CO},
       adsurl = {https://ui.adsabs.harvard.edu/abs/2026A&A...710A.113P}
}

@ARTICLE{Zhao_2022a,
       author = {{Zhao}, Xiaosheng and {Mao}, Yi and {Cheng}, Cheng and {Wandelt}, Benjamin D.},
        title = "{Simulation-based Inference of Reionization Parameters from 3D Tomographic 21 cm Light-cone Images}",
      journal = {\apj},
         year = 2022,
        month = feb,
       volume = {926},
       number = {2},
          eid = {151},
        pages = {151},
          doi = {10.3847/1538-4357/ac457d},
archivePrefix = {arXiv},
       eprint = {2105.03344},
 primaryClass = {astro-ph.CO},
       adsurl = {https://ui.adsabs.harvard.edu/abs/2022ApJ...926..151Z}
}

@INPROCEEDINGS{Makinen_2024,
       author = {{Makinen}, T. Lucas and {Sui}, Ce and {Wandelt}, Benjamin D. and {Porqueres}, Natalia and {Heavens}, Alan},
        title = "{Hybrid Summary Statistics}",
    booktitle = {Machine Learning and the Physical Sciences Workshop, NeurIPS 2024},
         year = 2024,
         note = {arXiv:2410.07548},
archivePrefix = {arXiv},
       eprint = {2410.07548},
 primaryClass = {stat.ML}
}

@ARTICLE{Zahn_2011,
       author = {{Zahn}, Oliver and {Mesinger}, Andrei and {McQuinn}, Matthew and {Trac}, Hy and {Cen}, Renyue and {Hernquist}, Lars E.},
        title = "{Comparison of reionization models: radiative transfer simulations and approximate, seminumeric models}",
      journal = {\mnras},
         year = 2011,
        month = jun,
       volume = {414},
       number = {1},
        pages = {727-738},
          doi = {10.1111/j.1365-2966.2011.18439.x},
archivePrefix = {arXiv},
       eprint = {1003.3455},
 primaryClass = {astro-ph.CO},
       adsurl = {https://ui.adsabs.harvard.edu/abs/2011MNRAS.414..727Z}
}

@ARTICLE{Greig_2018_21cmmc,
       author = {{Greig}, Bradley and {Mesinger}, Andrei},
      journal = {\mnras},
         year = 2018,
        month = jul,
       volume = {477},
       number = {3},
        pages = {3217-3229},
          doi = {10.1093/mnras/sty796},
archivePrefix = {arXiv},
       eprint = {1801.01592},
 primaryClass = {astro-ph.CO},
       adsurl = {https://ui.adsabs.harvard.edu/abs/2018MNRAS.477.3217G}
}

@ARTICLE{Baek_2009,
       author = {{Baek}, S. and {Di Matteo}, P. and {Semelin}, B. and {Combes}, F. and {Revaz}, Y.},
        title = "{The simulated 21 cm signal during the epoch of reionization: full modeling of the Ly-{\ensuremath{\alpha}} pumping}",
      journal = {\aap},
         year = 2009,
        month = feb,
       volume = {495},
       number = {2},
        pages = {389-405},
          doi = {10.1051/0004-6361:200810757},
archivePrefix = {arXiv},
       eprint = {0808.0925},
 primaryClass = {astro-ph},
       adsurl = {https://ui.adsabs.harvard.edu/abs/2009A&A...495..389B}
}

@ARTICLE{Semelin_2016,
       author = {{Semelin}, B.},
        title = "{Detailed modelling of the 21-cm forest}",
      journal = {\mnras},
         year = 2016,
        month = jan,
       volume = {455},
       number = {1},
        pages = {962-973},
          doi = {10.1093/mnras/stv2312},
archivePrefix = {arXiv},
       eprint = {1510.02296},
 primaryClass = {astro-ph.CO},
       adsurl = {https://ui.adsabs.harvard.edu/abs/2016MNRAS.455..962S}
}

@ARTICLE{Semelin_2007,
       author = {{Semelin}, B. and {Combes}, F. and {Baek}, S.},
        title = "{Lyman-alpha radiative transfer during the epoch of reionization: contribution to 21-cm signal fluctuations}",
      journal = {\aap},
         year = 2007,
        month = nov,
       volume = {474},
       number = {2},
        pages = {365-374},
          doi = {10.1051/0004-6361:20077965},
archivePrefix = {arXiv},
       eprint = {0707.2483},
 primaryClass = {astro-ph},
       adsurl = {https://ui.adsabs.harvard.edu/abs/2007A&A...474..365S}
}

@ARTICLE{Semelin_2017,
       author = {{Semelin}, B. and {Eames}, E. and {Bolgar}, F. and {Caillat}, M.},
        title = "{21SSD: a public data base of simulated 21-cm signals from the epoch of reionization}",
      journal = {\mnras},
         year = 2017,
        month = dec,
       volume = {472},
       number = {4},
        pages = {4508-4520},
          doi = {10.1093/mnras/stx2274},
archivePrefix = {arXiv},
       eprint = {1707.02073},
 primaryClass = {astro-ph.CO},
       adsurl = {https://ui.adsabs.harvard.edu/abs/2017MNRAS.472.4508S}
}

@inproceedings{Littwin_2024,
title={How {JEPA} Avoids Noisy Features: The Implicit Bias of Deep Linear Self Distillation Networks},
author={Etai Littwin and Omid Saremi and Madhu Advani and Vimal Thilak and Preetum Nakkiran and Chen Huang and Joshua M. Susskind},
booktitle={The Thirty-eighth Annual Conference on Neural Information Processing Systems},
year={2024},
url={https://openreview.net/forum?id=ez7w0Ss4g9}
}

@inproceedings{Balestriero_2024,
title={How Learning by Reconstruction Produces Uninformative Features For Perception},
author={Randall Balestriero and Yann LeCun},
booktitle={Forty-first International Conference on Machine Learning},
year={2024},
url={https://openreview.net/forum?id=XsDWw1Mn2p}
}

@InProceedings{He_2022,
    author    = {He, Kaiming and Chen, Xinlei and Xie, Saining and Li, Yanghao and Doll\'ar, Piotr and Girshick, Ross},
    title     = {Masked Autoencoders Are Scalable Vision Learners},
    booktitle = {Proceedings of the IEEE/CVF Conference on Computer Vision and Pattern Recognition (CVPR)},
    month     = {June},
    year      = {2022},
    pages     = {16000-16009}
}

\begin{appendix}
\nolinenumbers

\section{Pretraining loss curves}
\label{app:losses}
\begin{figure*}[h]
    \centering
    \includegraphics[width=\linewidth]{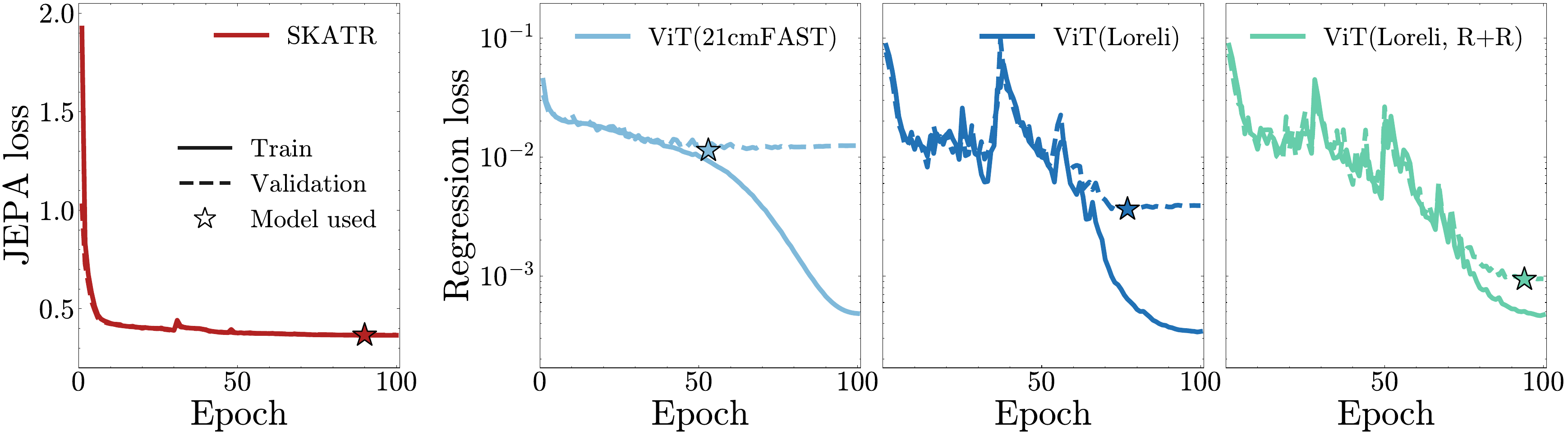}
    \caption{Training (solid) and validation (dashed) loss for the four encoders used at the noiseless full budget ($N=6517$). Stars mark the best-validation epoch, which is the checkpoint frozen and used for all downstream results. SKATR is shown on a linear axis in its own units (JEPA L1 loss in embedding space); the three supervised backbones share a logarithmic axis (L2 parameter-regression loss). The two objectives are not comparable in absolute value. SKATR shows no appreciable gap between training and validation at any epoch, whereas the supervised backbones separate after their best epoch; R+R augmentation visibly reduces that separation.}
    \label{fig:loss-curves}
\end{figure*}

Figure~\ref{fig:loss-curves} shows the training and validation loss of all four encoders behind the noiseless full-budget ($N=6517$) results. For every run the checkpoint used is the one with the lowest validation loss, not the final-epoch model; the epoch at which it was taken is marked by a star.

The self-supervised and supervised objectives are not on a common scale. SKATR minimizes an L1 loss between predicted and target embeddings (Sect.~\ref{sec:skatr}), whereas the supervised backbones minimize an L2 loss on the simulation parameters (Sect.~\ref{sec:baselines}); the two differ by orders of magnitude and measure different quantities.

SKATR's validation loss falls steeply over the first few epochs and is flat thereafter. Its best epoch is 90 of 100, at which point the validation loss is $0.3661$; over the final ten epochs the validation loss changes only by $0.03\%$. The run is thus well converged and shows no sign of overfitting.

The supervised backbones behave differently. Each reaches its best validation loss before the end of training -- epoch 53 for ViT(21cmFAST), epoch 77 for ViT(Loreli) -- after which the training loss continues to fall while the validation loss flattens. At their best epochs the ratio of validation to training loss is $1.2$ for ViT(21cmFAST) and $5.7$ for ViT(Loreli), growing to $25.7$ and $11.4$, respectively, by the final epoch. Since each encoder is frozen at its best epoch, this late-stage divergence does not propagate into the reported results, but it does show that the supervised objectives begin to fit features specific to their training set once the budget is large enough to allow it.

Adding R+R augmentation to ViT(Loreli) (Appendix~\ref{app:aug}) acts as a regularizer in exactly the expected way: the gap between training and validation loss narrows from a factor of $11.4$ to $2.0$ at the final epoch, and the best epoch moves from 77 to 94, so the model continues to improve on held-out data for almost the entire run. This supports the interpretation of Appendix~\ref{app:aug} that what R+R supplies to the supervised baseline is regularization rather than additional information.

\section{Marginal calibration: Rank statistics}
\label{app:sbc}

\begin{figure*}[h]
    \centering
    \includegraphics[width=\linewidth]{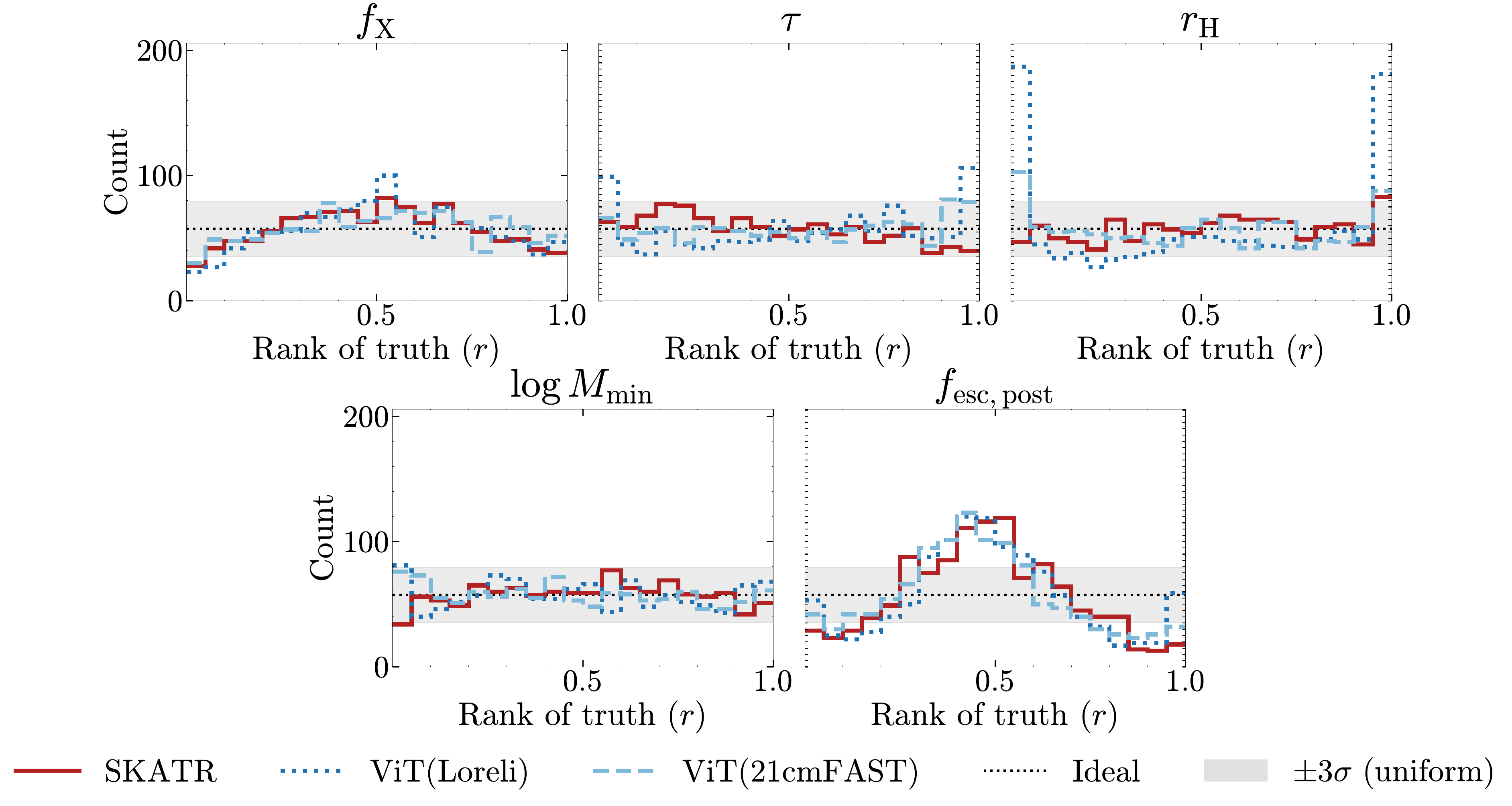}
    \caption{Per-parameter rank statistics (SBC) at the full Loreli budget
             ($N=6517$), noiseless. SKATR and ViT(21cmFAST) are flat on
             four of the five parameters and mildly underconfident
             ($\cap$-shaped) on $f_\mathrm{esc,post}$; ViT(Loreli) shares
             this $f_\mathrm{esc,post}$ underconfidence and is in addition
             visibly overconfident ($\cup$-shaped) on $r_\mathrm{H}$.}
    \label{fig:sbc-clean}
\end{figure*}

\begin{figure*}[h]
    \centering
    \includegraphics[width=\linewidth]{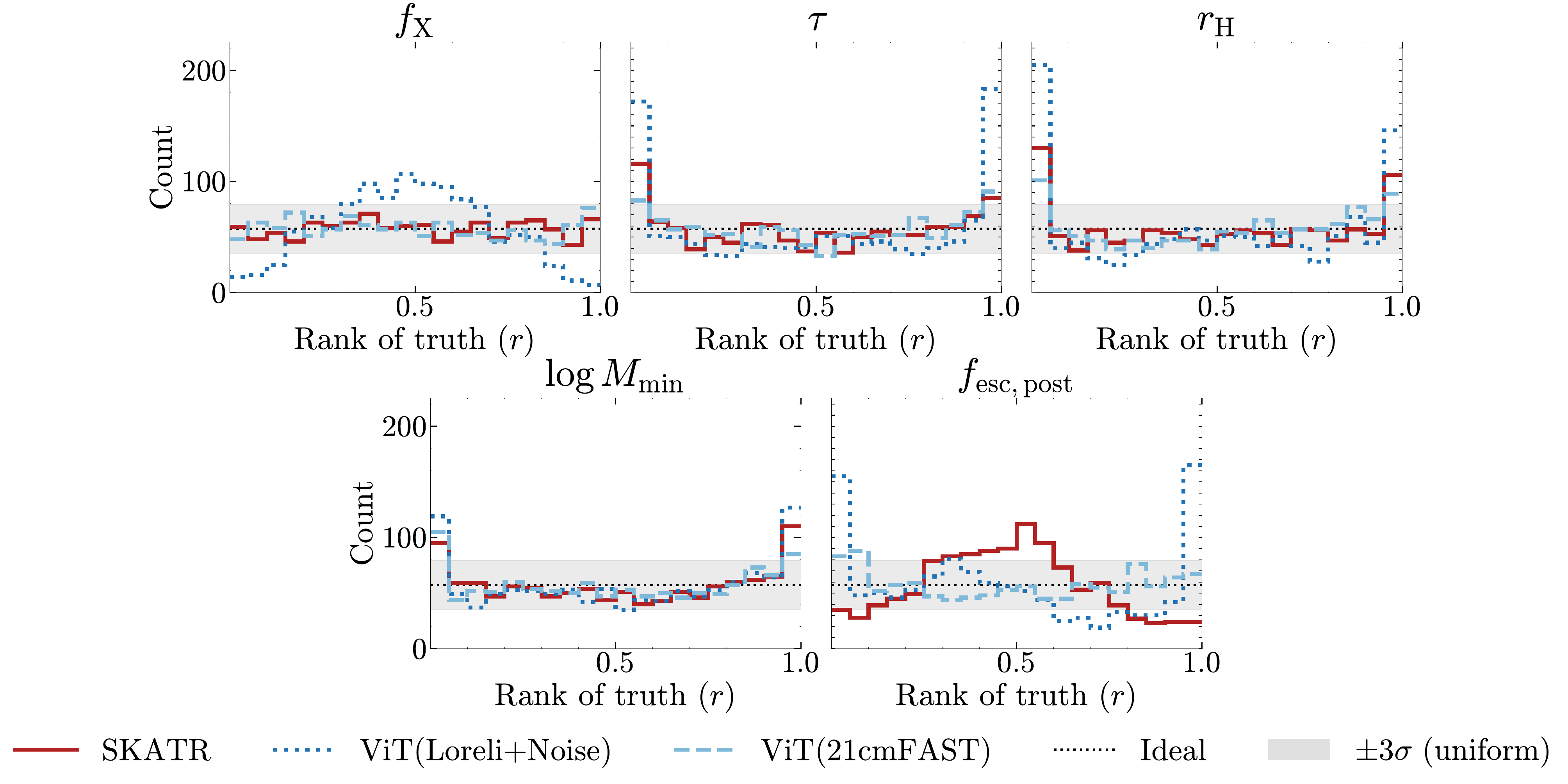}
    \caption{Same as Fig.~\ref{fig:sbc-clean} but under SKA AA$^\ast$ noise.
             SKATR is essentially flat on every parameter (only a small
             hint of overconfidence on $r_\mathrm{H}$); the noise-trained
             ViT(Loreli) is $\cup$-shaped on several parameters,
             matching the joint-TARP overconfidence of
             Sect.~\ref{sec:noisy-posteriors}; the zero-shot
             ViT(21cmFAST) shows indications of under- or overconfidence on several parameters.}
    \label{fig:sbc-noisy}
\end{figure*}

We complement the joint TARP coverage of Sect.~\ref{sec:posteriors}
and~\ref{sec:noisy-posteriors} with per-parameter simulation-based
calibration~\citep[SBC;][]{Talts_2018}. SBC tests, for each parameter
separately, whether the inferred posterior is statistically consistent
with the implicit prior-simulator distribution: for every test
light cone $i$ and parameter $p$ we record the rank of the true value
$\theta^{(i)}_p$ among the $M\times 500 = 2500$ samples drawn from the
combined-ensemble posterior, defined as the fraction of samples below
the truth. Over the held-out set of $1{,}150$ test light cones a
correctly calibrated marginal posterior produces ranks that are
uniformly distributed on $[0,1]$; the shape of the empirical rank
histogram hence provides a direct, low-variance diagnostic of the type of
miscalibration. A $\cup$-shape (mass piling up at the endpoints)
indicates posteriors that are too narrow, i.e.,\ overconfident; a $\cap$-shape (mass piling up in the center)
indicates posteriors that are too wide, i.e.,\ underconfident, which is
the more conservative failure mode; a monotonic tilt indicates a residual mean bias on
that parameter. In Figs.~\ref{fig:sbc-clean} and~\ref{fig:sbc-noisy}
the gray band marks the $3\sigma$ binomial spread expected under exact
uniformity given our test-set size and bin count.

The per-parameter picture is broadly consistent with the joint TARP shown in Figs.~\ref{fig:posterior-and-calibration} and~\ref{fig:posterior-and-calibration-noisy}. In the noiseless regime (Fig.~\ref{fig:sbc-clean}), SKATR and ViT(21cmFAST) are flat (well calibrated) on four of the five parameters and mildly $\cap$-shaped on $f_\mathrm{esc,post}$; ViT(Loreli) shares the $f_\mathrm{esc,post}$
underconfidence and is in addition visibly $\cup$-shaped and thus overconfident on
$r_\mathrm{H}$. Under SKA AA$^\ast$ noise
(Fig.~\ref{fig:sbc-noisy}), SKATR's per-parameter ranks are
essentially uniform across all five parameters, with only a small hint
of overconfidence on $r_\mathrm{H}$; the noise-trained ViT(Loreli)
becomes $\cup$-shaped on several parameters at once
(systematic overconfidence), and the zero-shot ViT(21cmFAST) shows
indications of under- or overconfidence on several parameters.

\section{Augmentation and the limits of supervised tuning}
\label{app:aug}

The supervised baselines (see Sect.~\ref{sec:baselines}) are trained with the same
architecture, the same optimization protocol, and without sample-level
augmentations. This is a deliberate choice for the baseline comparison: it isolates the pretraining objective as the only
difference between SKATR and the supervised ViTs, and it lets us
attribute any gap in accuracy, informativeness, or calibration to the
objective itself rather than to per-pipeline tuning. The natural
question, however, is what happens to that gap once the supervised
baseline is given a reasonable amount of target-specific
regularization.

Here we test the effect of augmentations on the in-domain ViT(Loreli), and retrain it with
R+R augmentations applied at training time: the
dihedral group $D_4$ of $90^\circ$ rotations and reflections about the two
transverse axes of the light cone, with the identity included. R+R is
a physically natural choice for 21cm light cones: in the absence of
foreground residuals and beam systematics, the underlying signal is
statistically isotropic in the sky plane, and explicitly enforcing the
sky-plane symmetry group at training time is a standard form of
data augmentation. All other hyperparameters of the ViT(Loreli)
backbone are kept fixed; only the augmentation set changes. We compare
three pipelines: SKATR (frozen, no augmentation), ViT(Loreli) (no
augmentation, the main-text baseline), and the R+R-augmented
ViT(Loreli) introduced here.

\begin{figure*}[h]
    \centering
    \begin{subfigure}{0.7\textwidth}
        \centering
        \includegraphics[width=\linewidth]{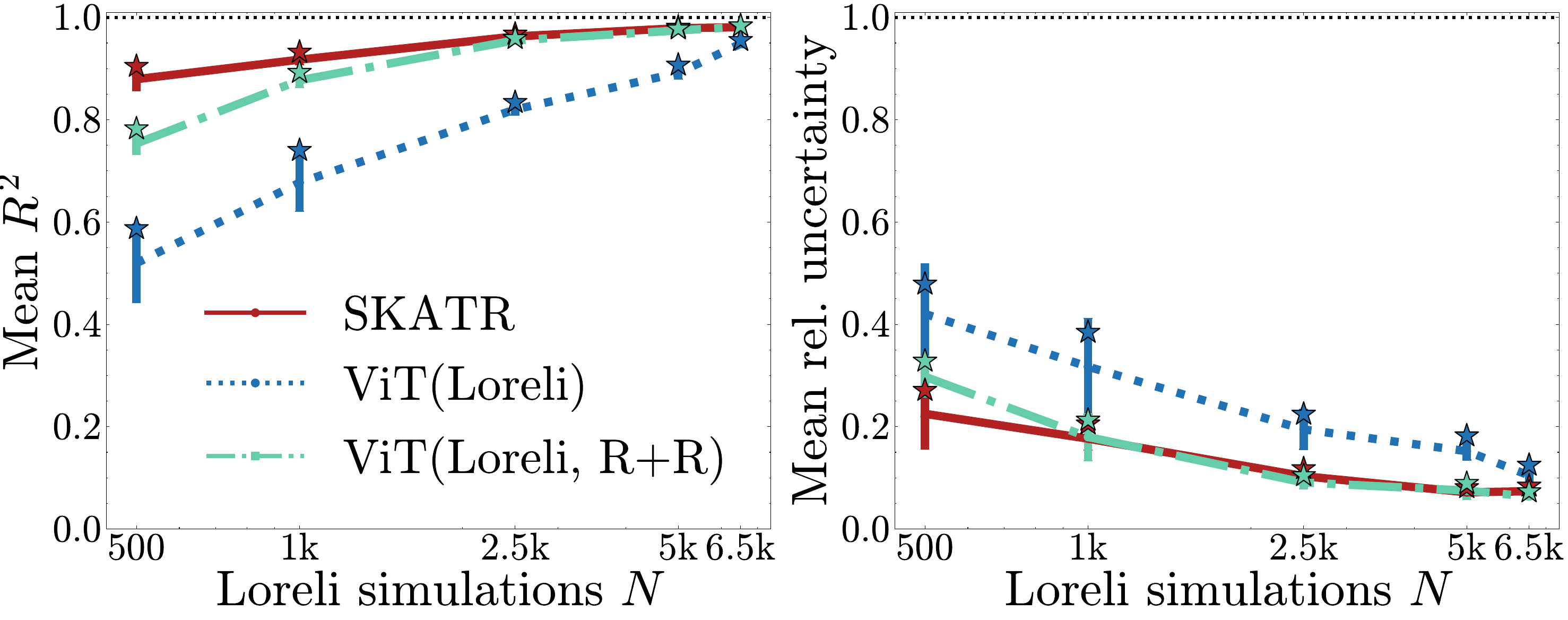}
        \caption{}
        \label{fig:scaling-appendix}
    \end{subfigure}
    \hfill
    \begin{subfigure}{0.29\textwidth}
        \centering
        \includegraphics[width=\linewidth]{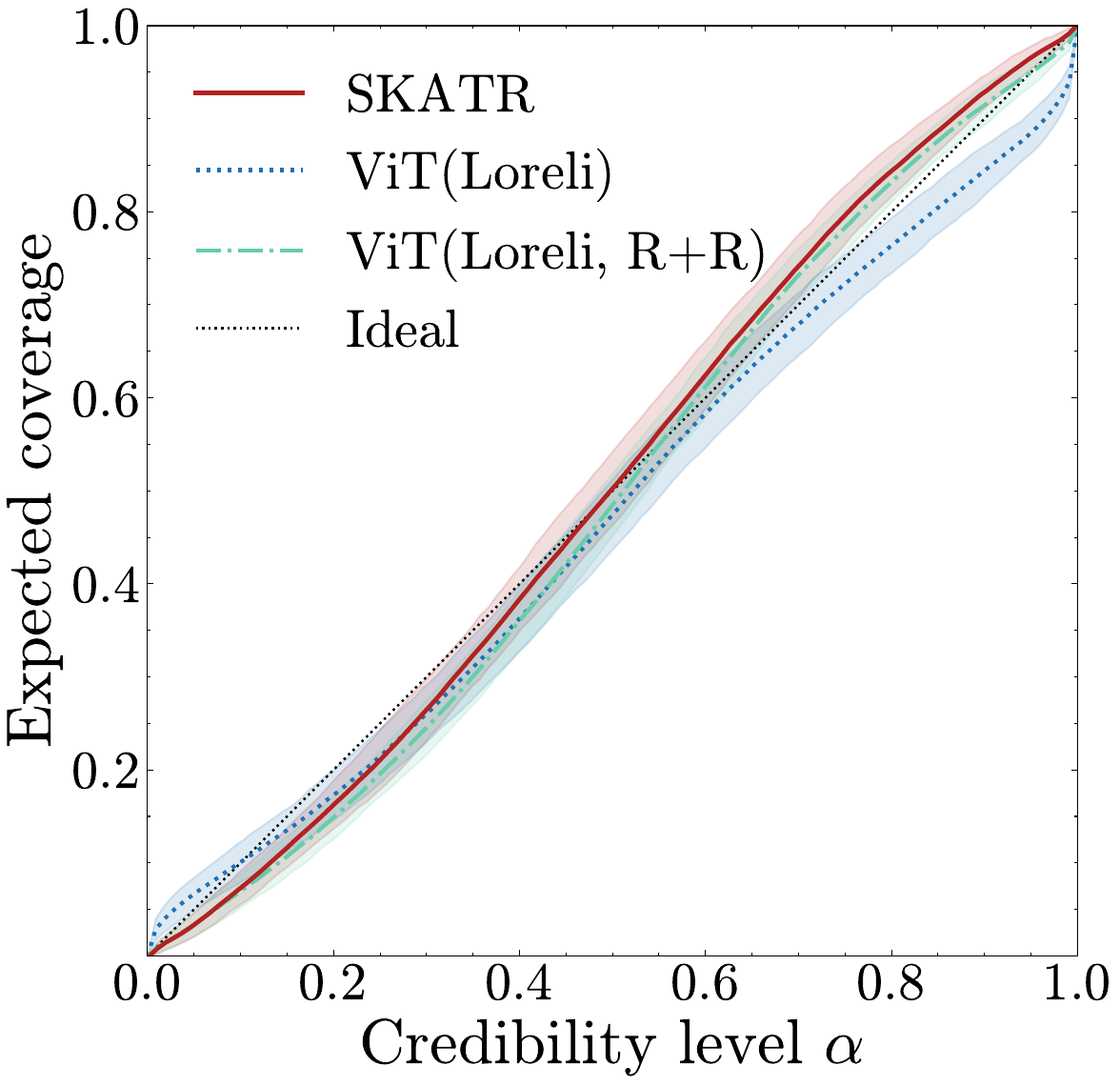}
        \caption{}
        \label{fig:tarp-appendix}
    \end{subfigure}
    \caption{R+R augmentation on the supervised ViT(Loreli) baseline,
             noiseless.
             (a) Accuracy and informativeness vs. the
             Loreli training budget $N$ for SKATR (no augmentation), ViT(Loreli)
             (no augmentation, the baseline from Sect.~\ref{sec:baselines}), and the R+R-augmented
             ViT(Loreli). The augmented baseline closes the accuracy
             and informativeness gap to SKATR by the full budget, and for known astrophysics and noise settings.
             (b) Joint TARP coverage at $N=6517$ for the same
             three pipelines: the R+R-augmented ViT(Loreli) now shows the same slight underconfidence as SKATR.}
    \label{fig:scaling-tarp-appendix}
\end{figure*}

\begin{figure*}[h]
    \centering
    \includegraphics[width=\linewidth]{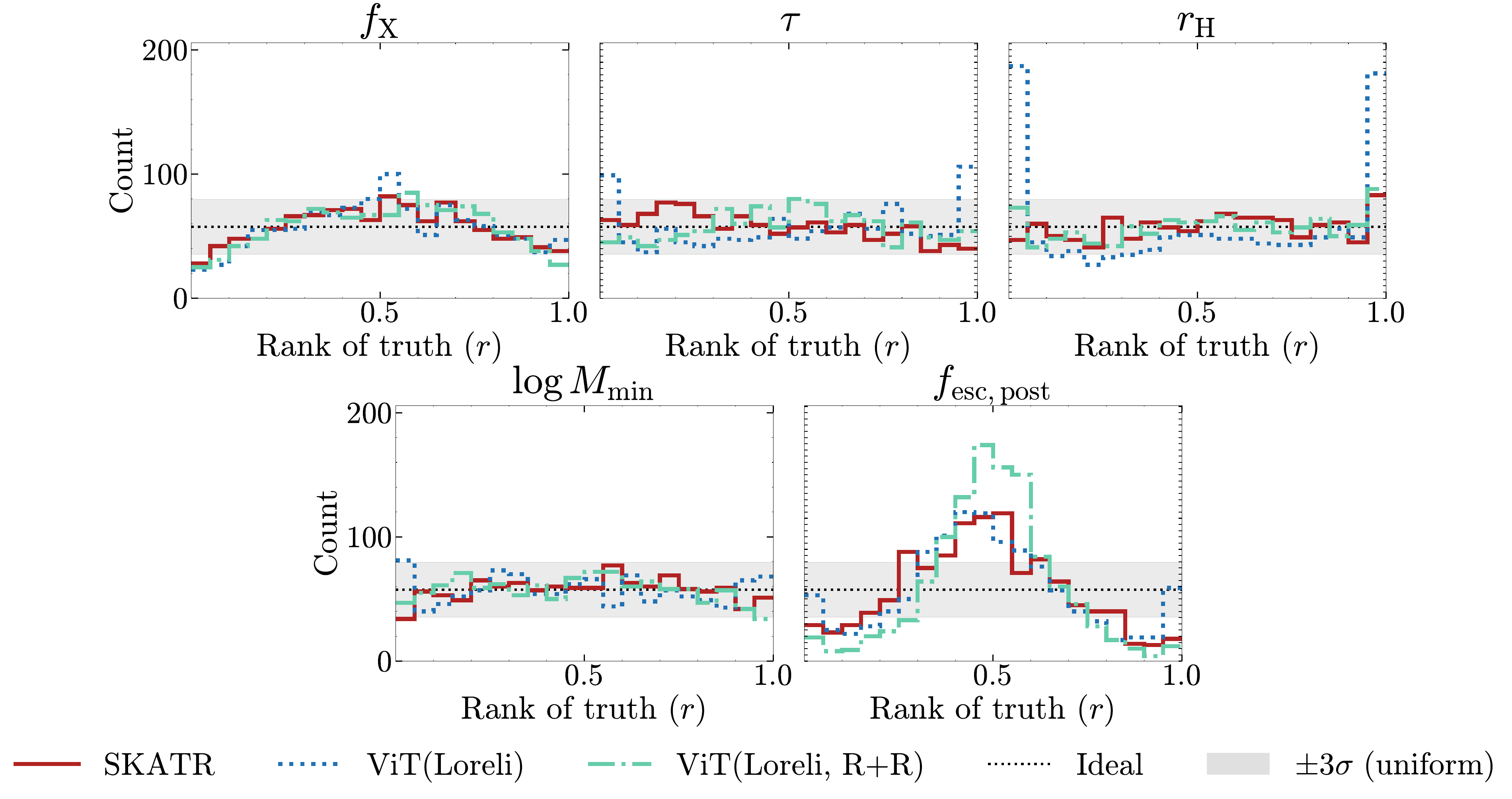}
    \caption{Per-parameter rank statistics (SBC) at the full Loreli
             budget ($N=6517$), noiseless, comparing SKATR (no augmentation),
             ViT(Loreli) (no augmentation), and the R+R-augmented
             ViT(Loreli). R+R augmentation removes the $\cup$-shape
             that the un-augmented ViT(Loreli) exhibits on $r_\mathrm{H}$,
             leaving the augmented baseline flat on every parameter except $f_\mathrm{esc,post}$,
             comparable to SKATR. The shared mild $\cap$-shape on
             $f_\mathrm{esc,post}$ is more pronounced for  ViT(Loreli) but persists across all three pipelines
             (see Sect.~\ref{sec:posteriors}).}
    \label{fig:sbc-appendix}
\end{figure*}

Figures~\ref{fig:scaling-appendix} and~\ref{fig:tarp-appendix} show that R+R narrows the accuracy, informativeness and calibration gap to SKATR depending on the training simulations budget and for fixed astrophysics and noise model. At the full budget ($N=6517$), the augmented ViT(Loreli) matches the SKATR mean $R^2$ to within statistical scatter and roughly matches (mean $0.07$ against $0.08$). The joint TARP (Fig.~\ref{fig:tarp-appendix}) of the augmented baseline indicates the same mild underconfidence as SKATR and the per-parameter rank histograms of Fig.~\ref{fig:sbc-appendix} show that the $\cup$-shape on $r_\mathrm{H}$ in the un-augmented ViT(Loreli) disappears under R+R, leaving the augmented baseline similarly flat as SKATR on every parameter except for $f_\mathrm{esc,post}$. Together, these results tell us that, with a reasonable augmentation choice, a fully supervised in-domain ViT can reach similar performance as SKATR for a sufficient number of expensive training simulations. Across the scaling sweep of Fig.~\ref{fig:scaling-appendix}, SKATR's lead in combined-ensemble mean $R^2$ narrows from $+0.04$ at $N=1000$ to $+0.01$ at $N=2500$ and vanishes at the full budget, but widens sharply once target labels become scarce: $+0.12$ at $N=500$. Sky-plane augmentations do not substitute for the pretraining of SKATR when fewer target labels are available. This is precisely the regime in which self-supervised pretraining on a low-cost source dataset is expected to pay off most: the encoder's physical inductive bias is set at pretraining time on $67{,}325$ 21cmFAST light cones rather than learned from the small target set, and the budget-$N$ Loreli labels are spent exclusively on the lightweight CFM head. The data-efficiency advantage is therefore complementary to, not subsumed by, the target-side augmentation route.

R+R is a physically motivated and broadly applicable augmentation for 21cm sky-plane analyses. Given enough target domain data a supervised pipeline equipped with R+R is a strong baseline, but it is not symmetry-agnostic: the dihedral group has to be identified as the relevant symmetry of the sky plane before training begins. The comparison shows that SKATR keeps pace with the strongest baseline tested here, while avoiding the several ways in which the supervised pipeline remains tied to the target data. First, a separate ViT(Loreli) encoder is trained from scratch at every budget $N$, so target labels are consumed twice (once to train the encoder, once to train the CFM head); SKATR reuses a single frozen encoder at every budget. Second, the regression objective is locked to the five Loreli parameters (Sect.~\ref{sec:loreli}) and would have to be redesigned for any other target simulator's parameterization; SKATR's encoder is parameter-set-agnostic by construction, and the parameters are seen only by the lightweight CFM head. Third, once realistic noise is added (Sect.~\ref{sec:results_noise}), a fresh supervised backbone has to be retrained from scratch on noisy data to recover accuracy, committing to a specific noise model at training time; SKATR transfers across the noiseless-to-mock shift zero-shot, without ever having seen the noise. For inference on actual SKA observations, where the underlying astrophysics and the precise noise model may differ from any training simulator, a summary that is not tied to the specifics of a particular simulator is the more robust foundation.

\end{appendix}

\end{document}